\documentclass[12pt,pdftex,noinfoline]{imsart}
\RequirePackage[OT1]{fontenc}
\usepackage{amsthm,amsbsy,amsmath,amsfonts,mathtools,amssymb}
\usepackage{natbib}
\usepackage{lipsum}
\usepackage{afterpage}

\usepackage{etoolbox}
\makeatletter
\patchcmd{\thebibliography}{\c@NAT@ctr\z@}{\c@NAT@ctr\z@
  \setlength{\labelwidth}{1.2em}%
  \setlength{\labelsep}{.5em}%
  \setlength{\leftmargin}{\dimexpr\labelwidth+\labelsep}%
}{}{}
\makeatother

\usepackage{mathrsfs}
\RequirePackage{hypernat}
\usepackage[ruled, vlined, lined, commentsnumbered]{algorithm2e}
\usepackage{graphicx}
\usepackage{verbatim}
\usepackage{times}
\usepackage{hyperref}
\usepackage[usenames,dvipsnames,svgnames,table]{xcolor}
\hypersetup{citecolor=MidnightBlue}
\hypersetup{linkcolor=MidnightBlue}
\hypersetup{urlcolor=MidnightBlue}
\usepackage{enumerate}
\usepackage{fullpage}
\usepackage[margin=1in,footskip=.40in]{geometry}
\usepackage{dsfont}
\usepackage[nameinlink,noabbrev]{cleveref}

\numberwithin{equation}{section}

\theoremstyle{remark}

\def\conditionA{Condition $A$}
\def\conditionAp{Condition $\tilde A$}
\def\conditionB{Condition $B$}

\def\ones{\mathds{1}}

\let\tilde\widetilde

\newcommand{\dd}[2]{{\frac{d#1}{d#2}}}

\newcommand{\Soft}{\text{Soft}}

\begin{document}

\setlength{\parskip}{0.5em}
\begin{frontmatter}
\runtitle{Neural Waves}
\title{Emergent organization of receptive fields in networks of excitatory and inhibitory neurons}
\begin{aug}
\vskip15pt
\address{
\begin{tabular}{c}
{\normalsize\rm\bfseries Leon Lufkin,\; Ashish Puri,\; Ganlin Song,\; Xinyi Zhong,\; John Lafferty}\\
\\[-7pt]
Department of Statistics and Data Science\\
Wu Tsai Institute \\[10pt]
Yale University
\end{tabular}
\\[10pt]
\today
\vskip10pt
}
\end{aug}

\begin{abstract}
Local patterns of excitation and inhibition that can generate neural waves are studied as a computational mechanism underlying the organization of neuronal tunings.
Sparse coding algorithms based on networks of excitatory and inhibitory neurons are proposed that exhibit topographic maps as the receptive fields are adapted to input stimuli. Motivated by a leaky integrate-and-fire model of neural waves, we propose an activation model that is more typical of artificial neural networks. Computational experiments with the activation model using both natural images and natural language text are presented. In the case of images, familiar ``pinwheel'' patterns of oriented edge detectors emerge; in the case of text, the resulting topographic maps exhibit a 2-dimensional representation of granular word semantics. Experiments with a synthetic model of somatosensory input are used to investigate how the network dynamics may affect plasticity of neuronal maps under changes to the inputs.
\end{abstract}

\end{frontmatter}


\section{Introduction}
\label{sec:intro}

Recordings of neurons in retina, visual cortex, and other brain regions reveal spontaneous activity that resembles propagating waves \citep{shadlen,han2008reverberation,santos2014radial,benucci2007standing,ackman2012retinal}.
The waves can appear in the absence of outside stimuli \citep{gribizis}, but can be disrupted in the presence of input \citep{churchland2010stimulus,sato2012traveling}, and affect sensitivity of perception \citep{muller2014stimulus,davis2020spontaneous,chemla2019suppressive}. While the mechanisms and roles of
neural waves are not fully understood, computational models
have shown that spontaneous traveling waves can result from local patterns of
inhibitory and excitatory recurrent connections, and that small changes to the relative balance of inhibition
and excitation can have large effects on cortical activity \citep{ermentrout2001traveling,keane,gong2009distributed,osan2001two,senk2020conditions,bressloff2000traveling,brunel2003determines,rubin,isaacson:2011}.

In this paper, local patterns of excitation and inhibition that can generate neural waves are studied as a computational mechanism underlying the organization of neuronal receptive fields. In early visual cortex (V1) in certain mammals, for example, the receptive fields are known to be edge detectors that are arranged topographically, with neighboring neurons coding for edges at similar scales and orientations \citep{bonhoeffer:1991,bonhoeffer:1993,crair,ackman2014role}.
In the leaky integrate-and-fire (LIF) model
proposed by \cite{keane}, neurons are arranged in a regular grid or randomly throughout a 2-dimensional sheet. Each neuron receives input from excitatory neurons within a local neighborhood, and from inhibitory neurons within another, typically larger neighborhood. The relative proportions of excitatory to inhibitory neurons and the sizes of the neighborhoods are design parameters to the model and influence the properties of the emergent wave patterns.
While previous investigations are focused on the wave characteristics as a function of network parameters \citep{brunel2000dynamics,keane}, here we study the emergent organization of the receptive fields when the neurons receive input stimuli.

We propose an abstraction of the leaky integrate-and-fire model as an activation model, where neurons do not have discrete firing events but rather
are associated with a continuous level of activation, as is typical of artificial neural
network models. The LIF and activation models use a similar local geometry of excitatory and inhibitory connections. The wave dynamics emerge in the activation model
through an iterative thresholding operator, which solves a constrained optimization problem. Under the activation model, the wave patterns saturate and converge, while under the LIF model the waves are dynamic and typically do not converge. Mathematically, the activation model is viewed as a regularization scheme based on a graph Laplacian having both negative and positive weights. When the local geometry is repeated regularly throughout a 2-dimensional grid, the application of the Laplacian to a vector of inputs is equivalent to a convolution operator. The scale of the wave fluctuations, in the absence of input stimuli, is seen in the eigenspectrum of the Laplacian. The regularization can also be viewed as a reaction-diffusion system \citep{turing,wooley,gilding}, but where
the reaction term appears as part of the diffusion operator, rather than as a separate potential.

We derive sparse coding algorithms based on this activation
model and find emergent patterns in the receptive fields
through a series of computational experiments, using both
natural images and natural text.
In the case of images, the resulting topographic maps resemble ``pinwheel'' patterns measured through neural recordings of the cat visual cortex \citep{bonhoeffer:1991,bonhoeffer:1993,crair,issa}. We also show how learning can take place in the LIF model, and report results with natural images in the supplement. We apply the same sparse coding algorithms used for images to distributed word representations obtained by training a language model on a large corpus of text. The resulting topographic maps exhibit a 2-dimensional representation of granular word semantics; the semantic units are thus analogous to visual edge detectors. We show through examples how the topographic map reveals elements of compositional semantics.

In a third computational experiment, we develop a model of somatosensory input.
Specifically, we experiment with a simplified model of a hand with five digits,
where neurons are tuned to input stimuli in an unsupervised manner.
When one of the ``fingers'' stops receiving input stimulus, the resulting region of the topographic map gradually fails to be activated. Notably, when the same recurrent geometry is used in an artificial neural network in a supervised fashion, the topographic map is not well formed. These findings can be compared against actual measurements of neural maps for the hand in sensory cortex \citep{makin,wesselink,huber} and other studies of motor control \citep{rubino2006propagating,zanos2015sensorimotor}.

The main contributions of this work are the proposed activation model, its formulation
as a convex optimization problem, associated algorithms for sparse coding, and experiments that demonstrate how this framework leads to emergent organization of receptive fields into topographic maps. The results presented here suggest that our framework for artificial neural networks has the potential to be used as a computational tool for better understanding the role of balanced excitation and inhibition in neuronal organization and plasticity, and as a biologically inspired building block for more complex neural network models.
This approach is in line with a growing body of work on biologically constrained computational models for machine learning and neuroscience \citep{lillicrap2016random,nokland2016direct,launay2020direct,frenkel2021learning,akrout,bellec,song,zhou21,ganguli19}.


\section{Motivation and previous work}
\label{sec:models}

Our starting point is the model of \cite{keane},
based on an $N \times N$ network of excitatory and inhibitory neurons whose dynamics are governed by a conductance-based integrate-and-fire model.
The voltage $V_{ij}(t)$ of the neuron at position $(i,j)$ in the grid evolves according to the differential equation
\begin{align}
  \label{eq:system}
    C \dd{}{t} V_{ij}(t)
    &= -g_L \bigl( V_{ij}(t) - V_L \bigr)
    -g_{ij}^E(t) \bigl( V_{ij}(t) - V_E \bigr)
    -g_{ij}^I(t) \bigl( V_{ij}(t) - V_I \bigr),
\end{align}
where $g_L$ is the ``leak'' conductance due to membrane resistance, and
$g^E$ and $g^I$ are conductances determined by the neighboring excitatory and inhibitory neurons.
Specifically, the excitatory and inhibitory conductances, $g_{ij}^E(t)$ and $g_{ij}^I(t)$, introduce a dependence on the firings of neighboring neurons according to
\begin{align}
    g^{E}_{ij}(t) &= F^{E} + \sum_{(i',j') \in E(i,j)} W_E \exp(-d^2(i,j;i',j')/2\sigma_E^2) \sum_l G^E(t-T^l_{i',j'}) \label{eq:ge}\\
    g^{I}_{ij}(t) &= F^{I} + \sum_{(i',j') \in I(i,j)} W_I \sum_l G^I(t-T^l_{i',j'}).\notag
  \end{align}
Here $E(i,j)$ is the set of excitatory neurons connected to the neuron at $(i,j)$ and $I(i,j)$ is the set of inhibitory neurons connected to this neuron; $T^l_{i,j}$ denotes the $l$-th firing time of the neuron.   Typical default parameters are $C = 1 \, \mu F$ for the capacitance, $g_L = 50 \, \mu S$ for the leak conductance, and $V_L = -70 \, mV$, $V_E = 0\ mV$, and $V_I = -80 \ mV$ as the reversal potentials for the leak, excitatory, and inhibitory voltages, respectively.

If the voltage $V_{ij}(t)$ reaches the threshold potential $V_T = -55 \,mV$, the neuron fires and its voltage is reset to the potential $V_R = -70 \,mV$ for a refractory period of $\tau_{ref} = 5 \,ms$,
in the default settings adopted in \cite{keane}. We fix $F^I = 2 \,\mu S$ and let $F^E$ vary throughout learning based on the similarity between the neuron's feedforward weights and the input stimulus, as described below. The kernels $G^E(t)$ and $G^I(t)$ describe the postsynaptic conductance of a neuron and are normalized to integrate to one.

\begin{figure}[t]
\begin{center}
\begin{tabular}{cccc}
\includegraphics[width=.22\textwidth]{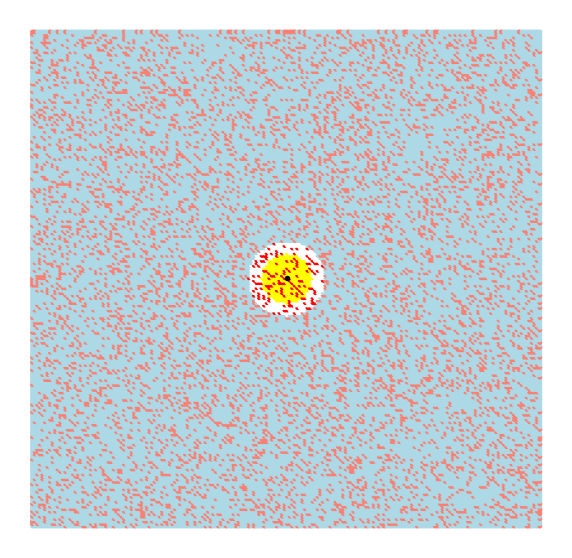}&
\includegraphics[width=.22\textwidth]{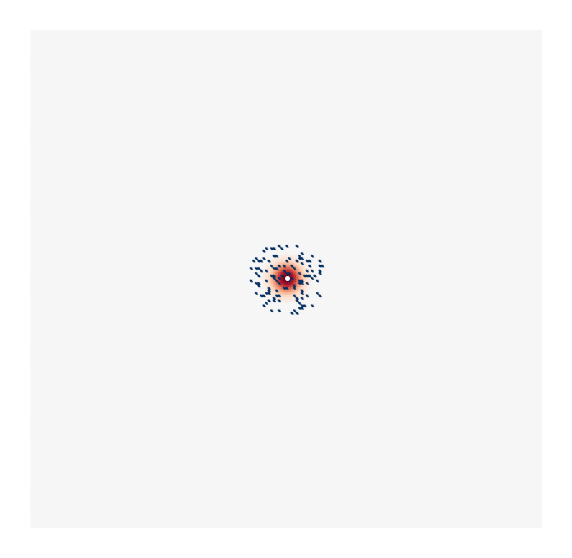}&
\includegraphics[width=.22\textwidth]{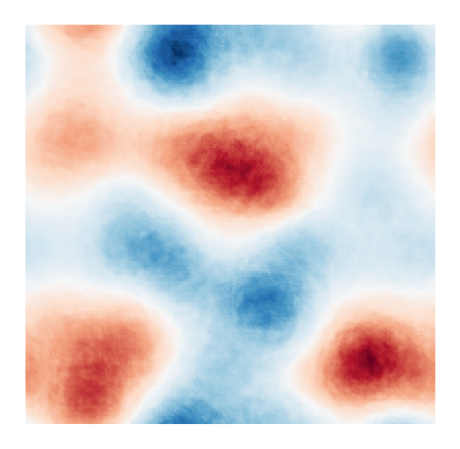}&
\includegraphics[width=.22\textwidth]{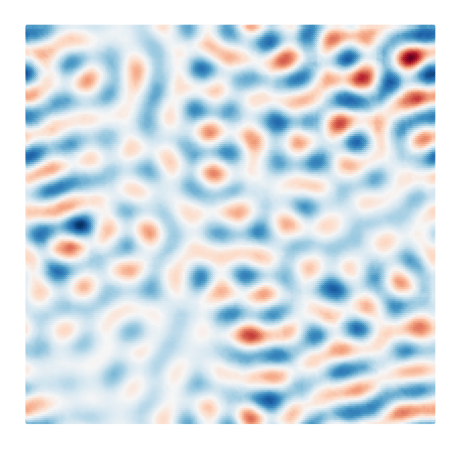}\\
\scriptsize{(a)} & \scriptsize{(b)} & \scriptsize{(c)} & \scriptsize{(d)}
\end{tabular}
\end{center}
\caption{Illustration of connectivity patterns in a grid of $200\times 200$ neurons.  Plot (a): In the neighborhood of the center neuron, the neurons within distance $d_E=10$ are shaded yellow, and the inhibitory neurons within distance $d_I=15$ are on a white background; inhibitory neurons colored red. Plot (b): excitatory neuron connections are weighted according to distance; the inhibitory neurons have a constant weight. Plots (c) and (d): eigenvectors of the Laplacian matrix of excitatory minus inhibitory weights, with periodic boundary conditions, showing the scaling of the wave patterns to be expected from the local connectivity.}
\label{fig:network}
\vskip-1pt
\end{figure}

While the focus of \cite{keane} is on understanding the emergence of propagating waves, here our goal is to study how the receptive fields of excitatory neurons might adapt to input stimuli. In the presence of an external stimulus
$X(t)$ the conductance of the excitatory term changes to
$  F^E + \Phi_{ij}^T  X(t) $
where $\Phi_{ij}$ denotes a vector of feedforward weights for the neuron at position $(i,j)$, if it is excitatory, acting as a linear filter. The system of differential equations \eqref{eq:system} defines the neural dynamics of the network; it is not directly tied to an objective function or loss for learning the filters $\Phi_{ij}$ from data. An example of the local geometry of excitation and inhibition we consider is given in Figure~\ref{fig:network}.


\section{An activation model based on wave dynamics}
\label{sec:activation}

The Euler dynamics of the differential equation \eqref{eq:system} is
given by the finite difference equations
\begin{align}
  \label{eq:euler}
     V_{ij}(t+dt)
    &= V_{ij}(t) -  \frac{dt}{C} \Bigl( g_L \bigl( V_{ij}(t) - V_L \bigr)
    +g_{ij}^E(t) \bigl( V_{ij}(t) - V_E \bigr)
    +g_{ij}^I(t) \bigl( V_{ij}(t) - V_I \bigr)\Bigr).
\end{align}
As the voltage approaches the firing threshold, the potential differences satisfy $V_{ij}(t)-V_E < 0$ with $V_{ij}(t) - V_I > 0$ and $V_{ij}(t) - V_L > 0$.
Thus, the excitatory term acts to increase the voltage, while the inhibitory and leak terms act to decrease the voltage. Our activation model abstracts the potential $V_{ij}$ and firing times $T^l_{ij}$ into a single scalar activation level $a_{ij}$. An analogue of the Euler step \eqref{eq:euler} is then the iterative update
\begin{align}
\label{eq:ista}
    a_{ij} \leftarrow \Soft_{\lambda} \Bigg( a_{ij} + \eta \bigg(\Phi_{ij}^TX-d_{ij}a_{ij}+
    \sum_{(i',j') \in E(i,j)} W_{ij;i'j'}^Ea_{i'j'}-\sum_{(i',j')\in I(i,j)} W_I a_{i'j'} \bigg) \Bigg)
\end{align}
for a small step size $\eta > 0$. Just as in the LIF model, $E(i,j)$ is the set of neurons within the excitatory radius $d_E$ and $I(i,j)$ is the set of neurons within the inhibitory radius $d_I$; The constant $W_I$ controls the strength of the inhibition, and the excitatory weights are given by
\begin{align*}
    W_{ij;i'j'}^E=W_E\exp(-(d^2(i,j;i',j'))/2\sigma_E^2).
\end{align*}
The constant $d_{ij} > 0$ incorporates the analogue of potential ``leak,'' and is specified below.
In place of a firing threshold, we incorporate the soft-thresholding
operator
\begin{equation*}
  \Soft_\lambda(u) = \mathop{\rm sign}(u) \max(|u|-\lambda, 0) =
  u \left(1 - \frac{\lambda}{|u|}\right)_+,
\end{equation*}
which can be thought of as a two-sided rectification.

It can now be seen that \eqref{eq:ista} is the
``iterative soft thresholding algorithm'' (ISTA) update for the optimization
\begin{equation}
  \label{eq:opt}
  \min_a \left\{ \frac{1}{2}\|\Phi^T X - a \|^2 + \frac{1}{2} a^T \Delta a  + \lambda \|a\|_1\right\}
\end{equation}
where the matrix $\Delta$ is a Laplacian matrix associated with
the excitatory and inhibitory weights. In vectorized form, the Laplacian is
\begin{equation*}
  \Delta_{i,j;i',j'}  =
  \begin{cases}
    d_{i,j}  & \text{if $(i,j) = (i',j')$} \\
    -W_{i,j;i',j'} & \text{otherwise}
  \end{cases}
\end{equation*}
where the graph weights are defined by
\begin{equation*}
  W_{i,j; i',j'} \equiv  W_{ij;i'j'}^E \ones ((i',j') \in E(i,j)) - W_I \ones ((i',j') \in I(i,j)).
\end{equation*}
The matrix $\Delta$ is positive semi-definite if it is diagonally dominant; that is, if
\begin{equation}
  \label{eq:diagdom}
  d_{i,j} \geq
  \sum_{(i',j') \in E(i,j)} W_{ij;i'j'}^E +  \sum_{(i',j') \in I(i,j)} W_I.
\end{equation}
Assuming condition \eqref{eq:diagdom}, the optimization \eqref{eq:opt} is convex,
and the set of minimizers is non-empty and convex. Moreover, for sufficiently small  step size $\eta$
the iterative algorithm \eqref{eq:ista} will converge to a minimizer \citep{beck:ista}.

\subsection{Properties of the optimization}

The convex optimization \eqref{eq:opt} has three terms which are in tension with one another, each of which contributes different properties to the solutions.

\noindent
{\it\bfseries Response to the stimulus.} The squared error term $\frac{1}{2}\|\Phi^T X - a \|^2$ encourages the activations to respond to the stimulus. For an excitatory neuron at grid position $(i,j)$, the inner product
$\Phi_{ij}^T X$ is large (in absolute value) if the neuron's feedforward weights $\Phi_{ij}$, acting as a linear filter,
matches the stimulus $X$.  In the absence of any other terms in the optimization, the closed-form solution is
\begin{equation*}
  a_{ij} = \Phi_{ij}^T X.
\end{equation*}

\noindent
{\it\bfseries Sparsity.} The $\ell_1$ penalty $\lambda \|a\|_1$ encourages sparse solutions. In the absence
of the Laplacian term, the optimization
\begin{equation*}
  \min_a \left\{ \frac{1}{2}\|\Phi^T X - a \|^2 + \lambda \|a\|_1\right\}
\end{equation*}
also has a closed-form solution, given by
\begin{align*}
    a_{ij} = \Soft_{\lambda} \left(\Phi_{ij}^T X\right).
\end{align*}
In this way, if the neuron's tuning only weakly matches the stimulus, with $|\Phi_{ij}^T X| \leq \lambda$ then
the activation would be zero, $a_{ij} = 0$, which corresponds to the voltage being below threshold in the firing model.

\noindent
{\it\bfseries Wave patterns.} The Laplacian regularization $\frac{1}{2} a^T \Delta a$ encourages wave patterns in the solutions. This can be seen in the right two plots of Figure~\ref{fig:network}, which correspond to two
eigenvectors of the Laplacian. Under condition \eqref{eq:diagdom}, we have that $a^T \Delta a \geq 0$.
The Laplacian regularization has three components, corresponding to excitation, inhibition, and ``leak."
The excitatory weights encourage neighboring neurons to ``fire together,'' in the sense that neighboring neurons will tend to have similar activations. The inhibitory terms discourage the activation from spreading, playing the role of a reaction term in a reaction-diffusion system. And the diagonal term corresponds to voltage leak; setting the diagonal sufficiently large ensures that the optimization is well defined.

The optimization incorporates all three terms---stimulus, sparsity, and regularization---and
the iterative soft thresholding algorithm \eqref{eq:ista} can be more succinctly  written as
\begin{equation*}
  a \leftarrow \Soft_{\lambda} \left(a + \eta \left(\Phi_{ij}^T X - \Delta a\right)\right).
\label{eq:ours}
\end{equation*}

Our derivation of the activation model comes from the Euler method for integrating the
LIF differential equations. The optimization \eqref{eq:opt} does not have
an intrinsic time scale; rather, a time scale is implied by the iterative soft thresholding
algorithm \eqref{eq:ista}, and corresponds to the time scale of the Euler method. However,
the soft-thresholding algorithm converges (under assumption \eqref{eq:diagdom}), while the
Euler iterates need not converge. This is due to the refractory period after a neuron fires,
which can lead to propagation of waves.

\noindent
{\it\bfseries Reconstruction error.}
The optimization \eqref{eq:opt} is not directly
targeting the error $\|X - \Phi a\|^2$ for reconstructing the stimulus from the activations. Under an $\ell_1$  sparsity penalty, the standard sparse coding procedure of \cite{Olshausen:Field:96}
is based on the lasso optimization
$
\min_a  \frac{1}{2} \|X - \Phi a\|^2 + \lambda \|a\|_1
$
with associated ISTA iterates
\begin{align}
  a_{ij} & \leftarrow \mbox{\rm Soft}_{\lambda}\left(a_{ij} + \eta
  \Bigl(\Phi_{ij}^T X - a_{ij} -
  \sum_{i',j' \neq i,j} a_{i'j'}\Phi_{ij}^T \Phi_{i'j'}
  \Bigr)\right),
\label{eq:theirs}
\end{align}
assuming the feedforward weights $\Phi$ are normalized so that $\|\Phi_{ij}\|=1$.
The last term acts as a lateral inhibition effect \citep{rozell:08},
but is non-local, coupling together the filters for neurons
that are far apart.
In a similar manner, \cite{gregor2010emergence} augment the reconstruction error
with a structured lateral inhibition term. The approach of \cite{hoyer} learns
topographic maps using a two-layer model of complex cells, based on
reconstruction error through a likelihood function. Comparing
\eqref{eq:ours} and \eqref{eq:theirs}, it is seen that our proposed activation
model replaces lateral inhibition based on feedforward weights
with lateral excitation and inhibition based on fixed, local connectivities.

\subsection{Tuning the feedforward weights and receptive fields}

When iteratively tuning the weights $\Phi$ as input stimuli
are received, we update the weights using the gradient of
the squared error objective:
\begin{align}
\label{eq:rf_update}
\Phi_{ij} & \leftarrow \Phi_{ij} + \gamma \sum_{k\in B} a^{(k)}_{ij} \left(X^{(k)} - \Phi a^{(k)}\right)
\end{align}
where $B$ is a mini-batch of randomly selected inputs.
Here $\gamma>0$ is a small step size, and $a^{(k)}_{ij}$ is the activation for the excitatory neuron
at location $(i,j)$ in the grid in response to the $k$th input in the mini-batch.

In each epoch of training, the activations over the mini-batch are iterated until convergence,
and then a gradient step is made using the above update.
A similar update can be used under the LIF model, where the
Euler steps are made for a fixed number of iterations and the firing counts are
used in the gradient step for the feedforward weights.
Since neither the activations nor the firings are
selected to minimize reconstruction error, this algorithm
is not direct stochastic gradient descent; but we observe
that it converges empirically.

We note that this update to the feedforward weights fields is non-local in the sense that the
change in weight at neuron $(i,j)$ depends on the global residual $X^{(k)} - \Phi a^{(k)}$ involving
all of the other activations. When the activations are very sparse, it is expected that
this is well approximated by
\begin{align}
\sum_{k\in B} a^{(k)}_{ij} \left(X^{(k)} - \Phi a^{(k)}\right)
& \approx \sum_{k\in B} a^{(k)}_{ij}\, X^{(k)}  - \Phi_{ij} \sum_{k\in B} a_{ij}^{(k)\,2}.
\end{align}
This is the local update rule that is
proposed by \cite{zylberberg}, and is a form of Oja's rule for Hebbian learning
\citep{oja}. A justification of this approximation is that if each neuron fires
with probability $p \ll 1$, then $\frac{1}{|B|}
\sum_{k\in B} a_{ij}^{(k)\,2} = O(p)$ while
$\frac{1}{|B|}\sum_{k\in B} a^{(k)}_{ij} a^{(k)}_{i'j'} = O(p^2)$ for neurons at locations $(i,j)$ and $(i',j')$ that fire roughly independently. However, this
local update rule is not effective when the neurons are organized topographically,
since neighboring neurons are strongly correlated.

The receptive field of a neuron depends on the neuron's feedforward weights as well as the responses of other neurons through lateral connections, both excitatory and inhibitory.  Conceptually, the receptive field for a neuron at position $(i,j)$ in the grid can be thought of as the derivative $\frac{\partial a_{ij}(X)}{\partial X}$.
To compute the receptive field we input delta functions having
a spike in a given pixel position, and zeros elsewhere, and then standardize
the stimulus to have mean zero and variance one.   The
receptive field for that pixel position is then the response
of the neuron to that stimulus \citep{Olshausen:Field:96}. The resulting
receptive fields are very similar to the feedforward weights, but
sharpened and more selective, due to the sparsity imposed in the activations.

\begin{figure*}[t!]
\begin{tabular}{cc}
\begin{tabular}{c}
\hskip-10pt
\includegraphics[width=.55\textwidth]{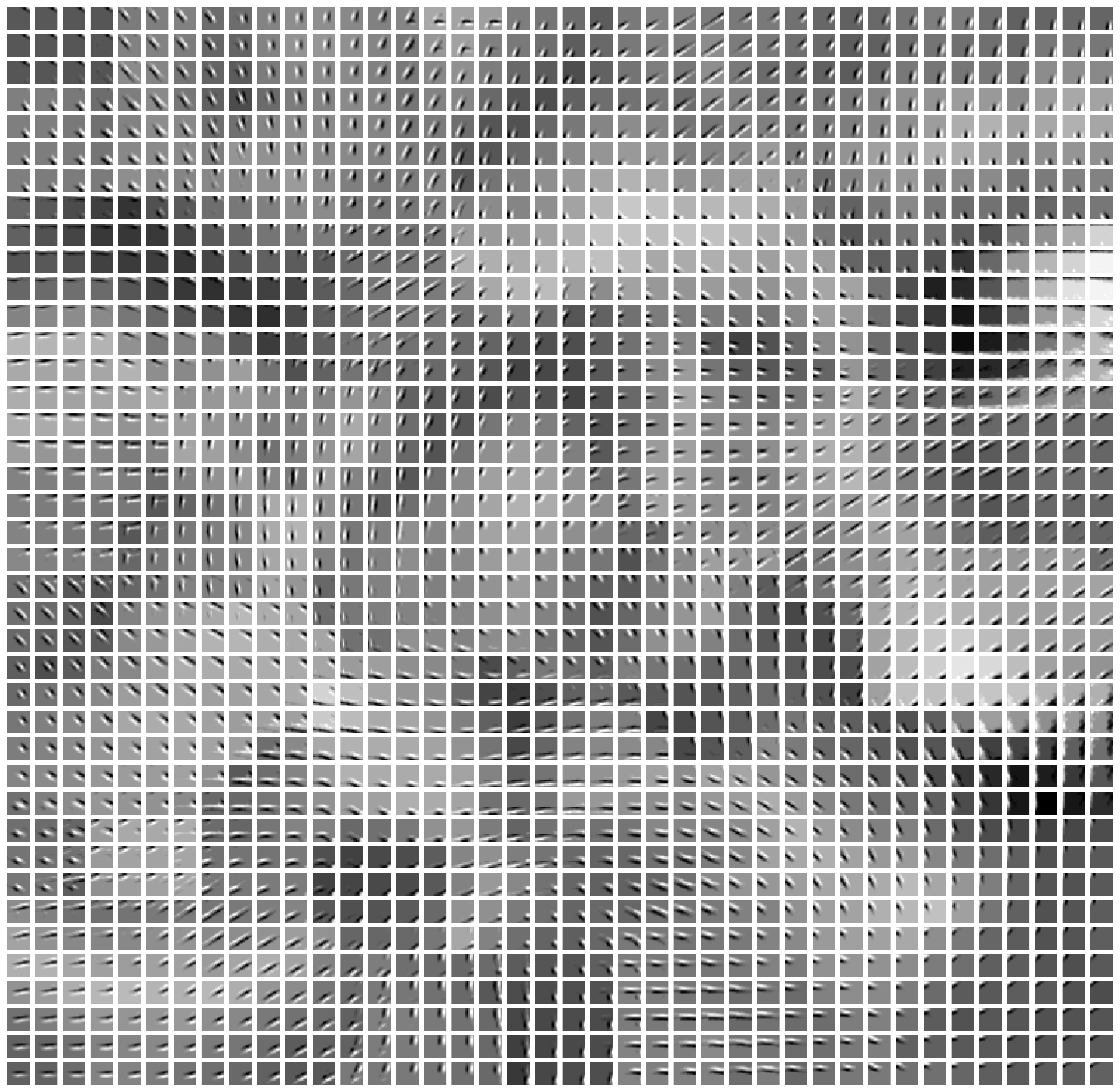}
\end{tabular}
&
\begin{tabular}{c}
\begin{tabular}{c}
\includegraphics[width=.30\textwidth]{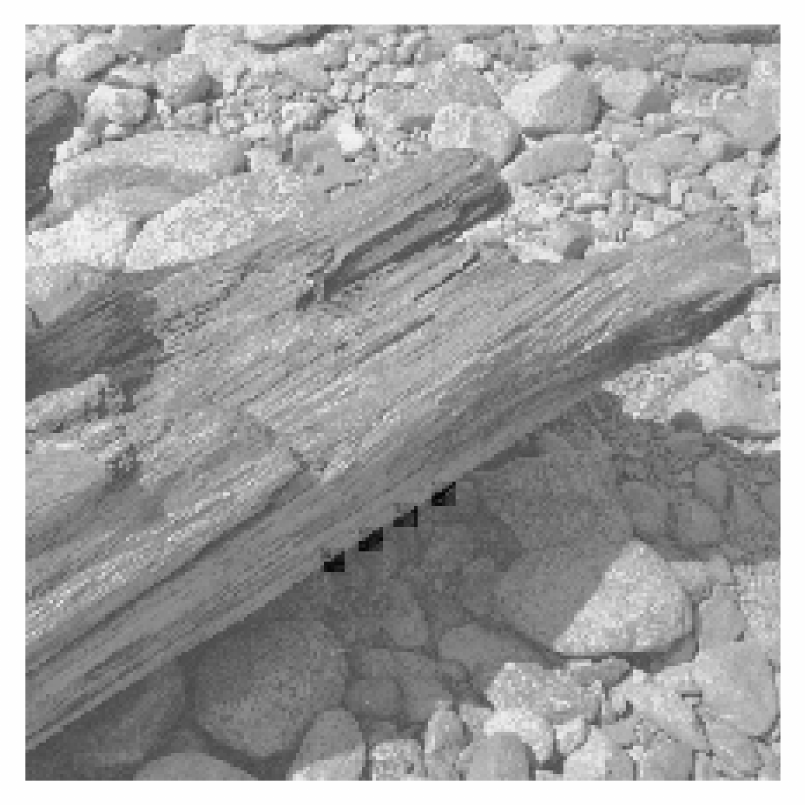}
\end{tabular}
\\
\begin{tabular}{cc}
\hskip-5pt
\raisebox{5pt}{\includegraphics[width=.06\textwidth]{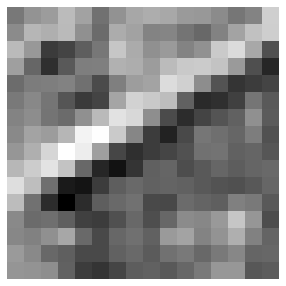}} \quad
\hskip-5pt\includegraphics[width=.12\textwidth]{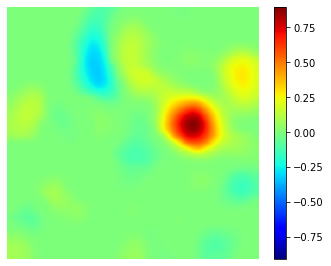}   &
\hskip-5pt
\raisebox{5pt}{\includegraphics[width=.06\textwidth]{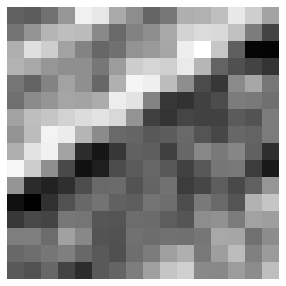}} \quad
\hskip-5pt\includegraphics[width=.12\textwidth]{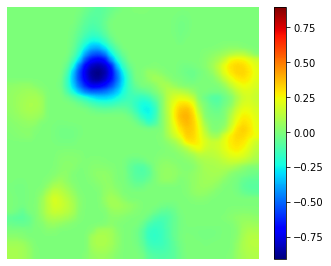} \\
\hskip-10pt
\raisebox{5pt}{\includegraphics[width=.06\textwidth]{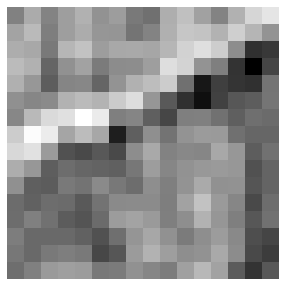}} \quad
\hskip-5pt\includegraphics[width=.12\textwidth]{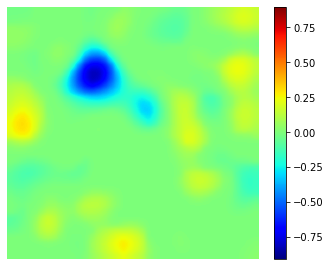}  &
\hskip-5pt
\raisebox{5pt}{\includegraphics[width=.06\textwidth]{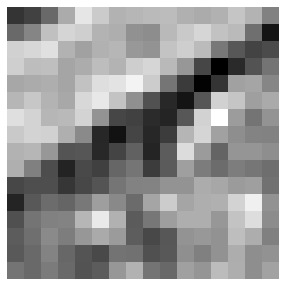}} \quad
\hskip-5pt\includegraphics[width=.12\textwidth]{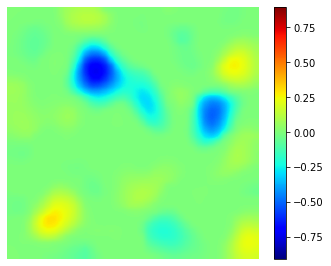}
\end{tabular}
\end{tabular}
\end{tabular}
\caption{Left: Receptive fields for a $40\times 40$ grid of excitatory neurons, coding for $16\times 16$ patches randomly drawn from natural images. The receptive fields
organize to form ``pinwheel'' patterns, with nearby neurons responding to similarly oriented edges, as has been measured in V1 experimentally \citep{crair}.
Right: Image patches and the resulting activations $[a_{ij}]$ of excitatory neurons;
plotted as heatmaps with Gaussian smoothing.}
\label{fig:imagecodewords}
\end{figure*}

As seen in the following experiments, the receptive fields under this model naturally exhibit topographic maps as the weights are tuned to input stimuli. In contrast to other machine learning frameworks for sparse coding
\citep{Olshausen:Field:96,Bell:97,rozell:08,zylberberg}, the activations are not
directly driven by reconstruction error. Previous work in machine learning for structured sparse coding of natural images is closely related \citep{gregor2010emergence,hoyer}.


\section{Organization of tunings to natural images}
\label{sec:images}

In this section we give examples of the topographic map
that results from running the algorithm described above. Our setup consists of two coupled $40\times 40$ grids of neurons; one grid is excitatory with each
neuron receiving an input stimulus, and one grid is inhibitory.
A  neuron at a given grid point is connected to all excitatory neurons within a radius $d_E=3$
and to all inhibitory neurons within a radius $d_I = 5$; the relative strengths
of the excitatory and inhibitory neurons were determined by the parameters $W_E=30$ and $W_I=5$
in equation \eqref{eq:ista}.
The receptive fields of the excitatory neurons are tuned using stimuli that are
random $16\times 16$ patches from natural images; we use the original
images from \cite{Olshausen:Field:96} which are publicly available
at {\small\url{http://www.rctn.org/bruno/sparsenet}}.
The algorithm was run for 10,000 epochs with mini-batch size 128, where image
patches are sampled randomly from the set of ten natural images. The patches are standardized to have
unit standard deviation for each pixel.
\afterpage{
\begin{figure}[t!]
\begin{tabular}{c}
\includegraphics[width=.30\textwidth]{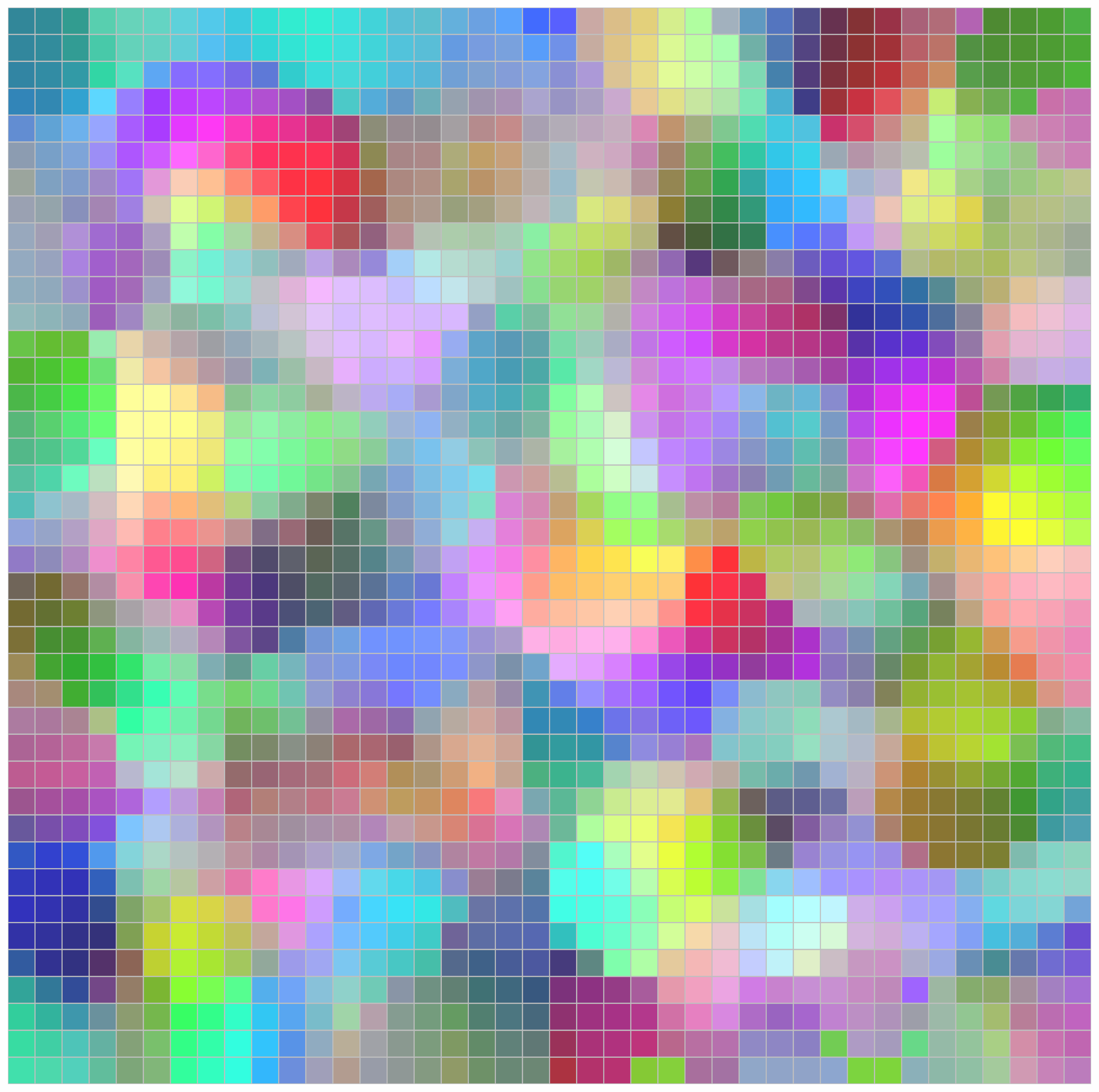}\\[10pt]
\includegraphics[width=.30\textwidth]{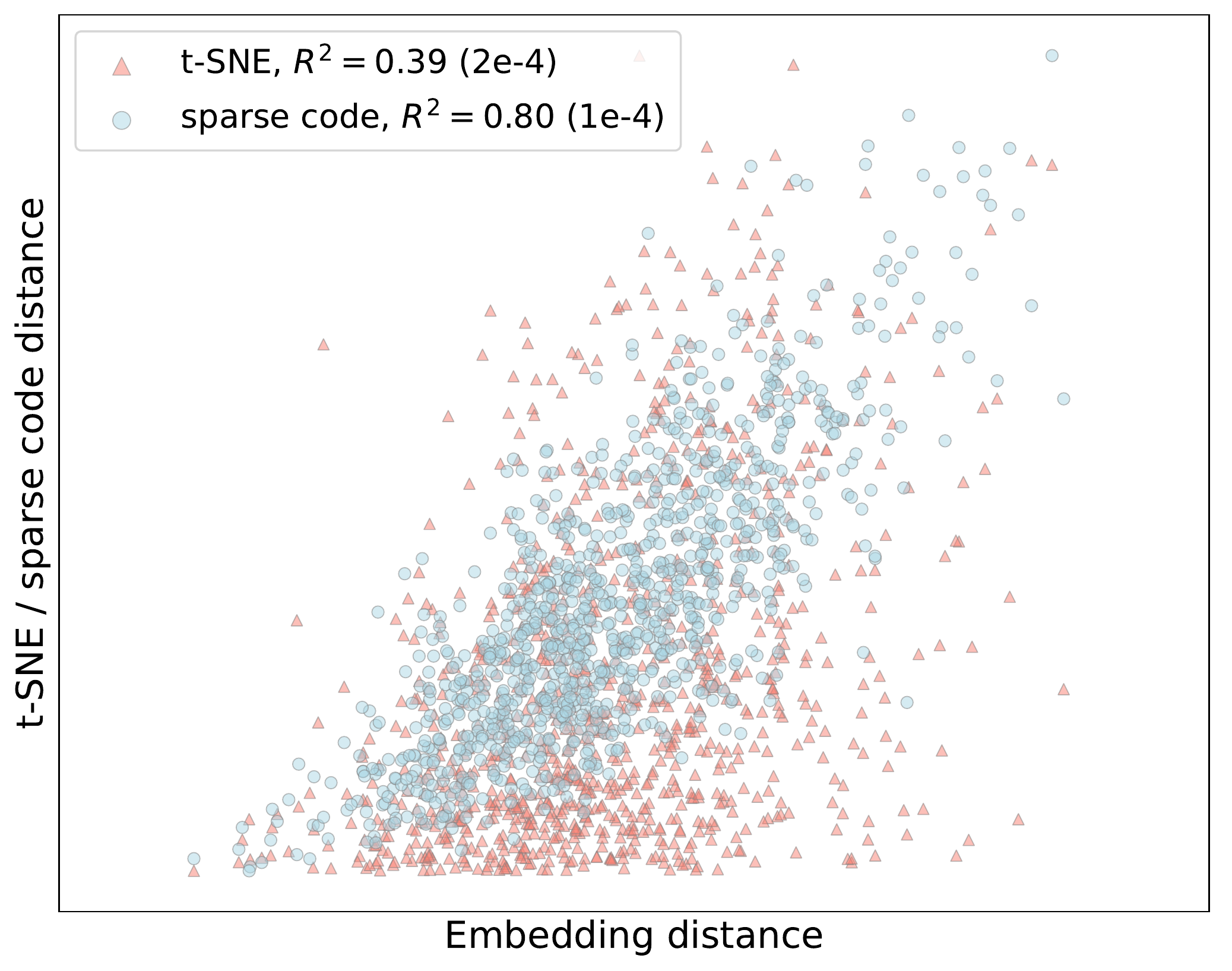}
\end{tabular}
\quad
\hskip-10pt
\begin{tabular}{c|c|c|c}
\hskip-5pt
\includegraphics[width=.14\textwidth]{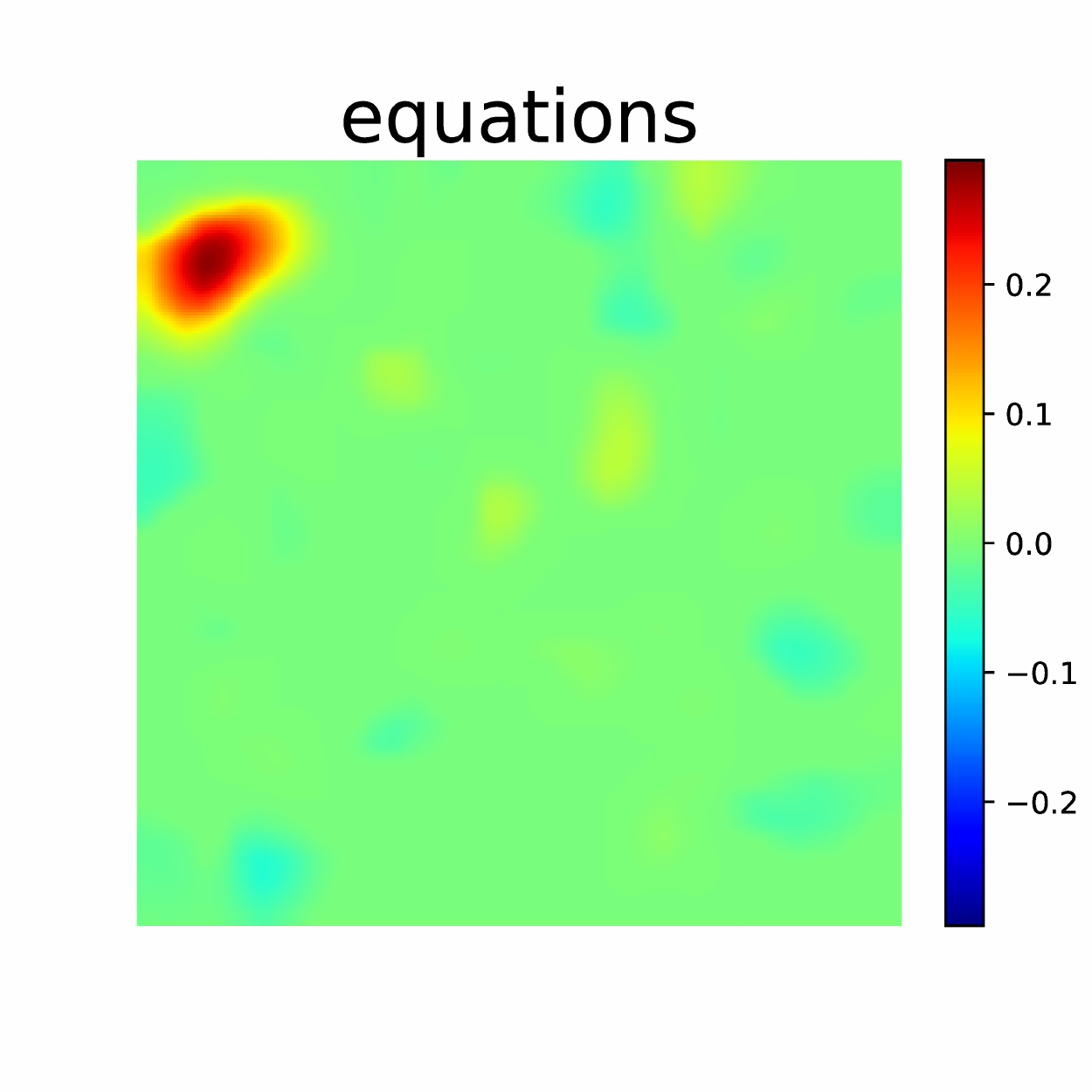} &
\includegraphics[width=.14\textwidth]{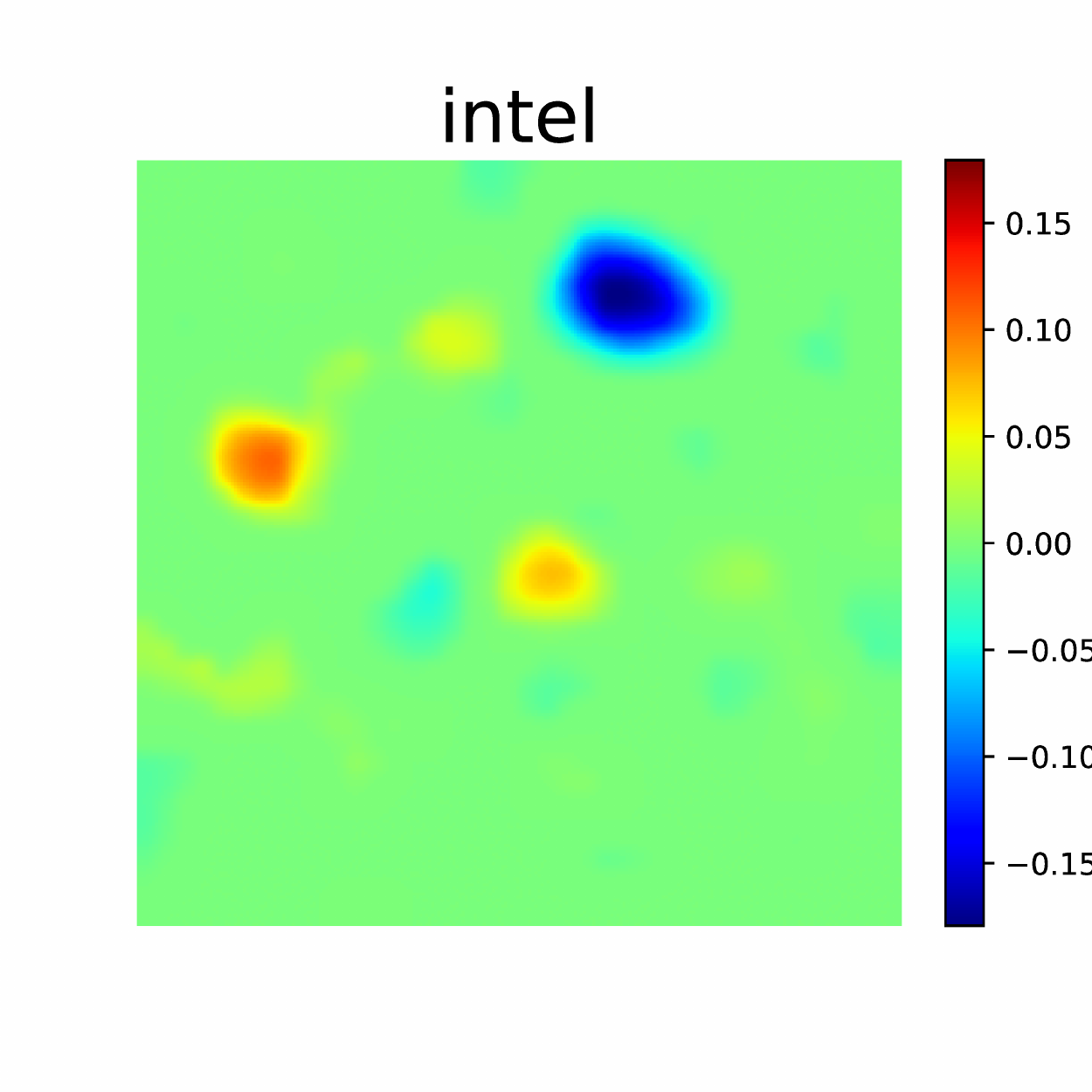} &
\includegraphics[width=.14\textwidth]{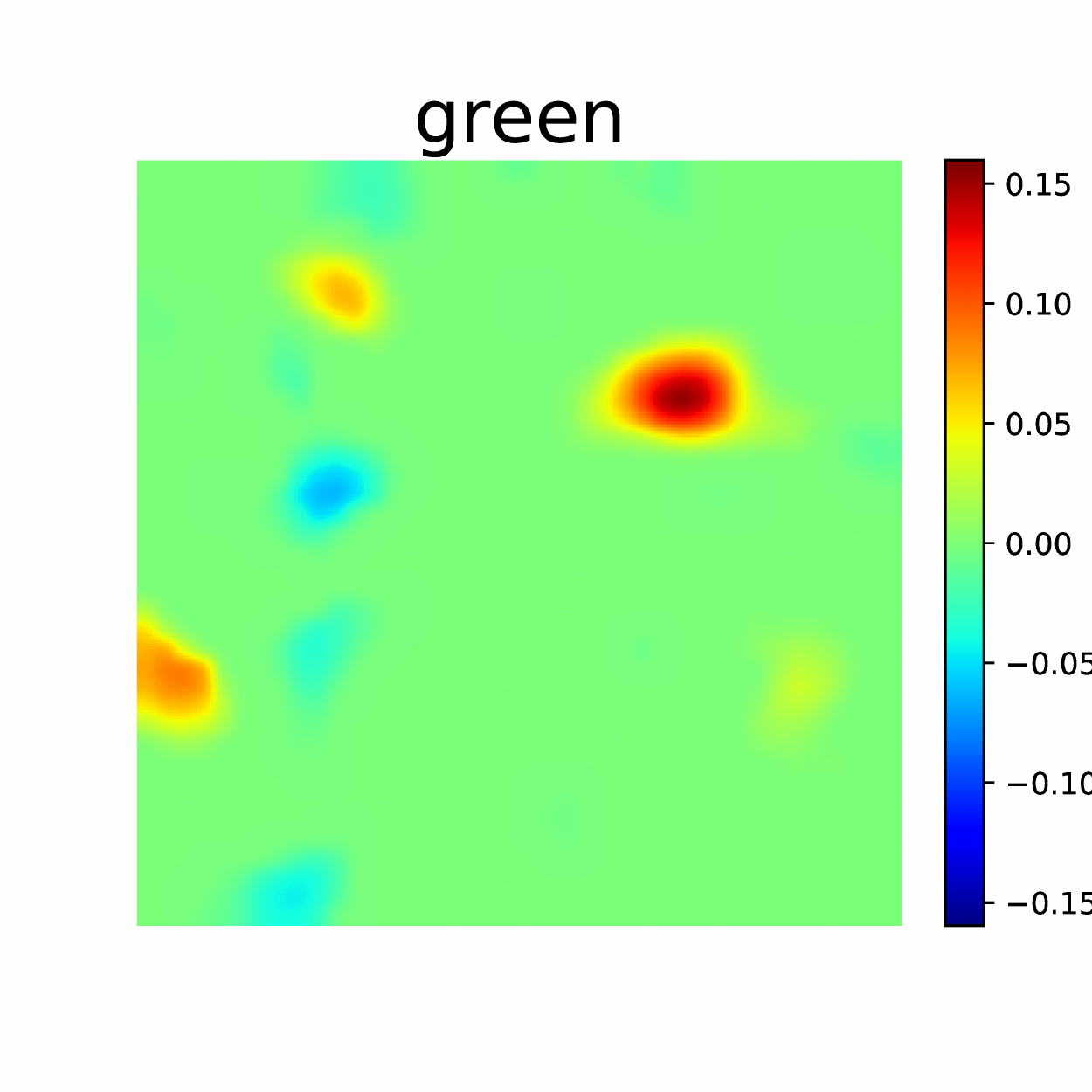} &
\includegraphics[width=.14\textwidth]{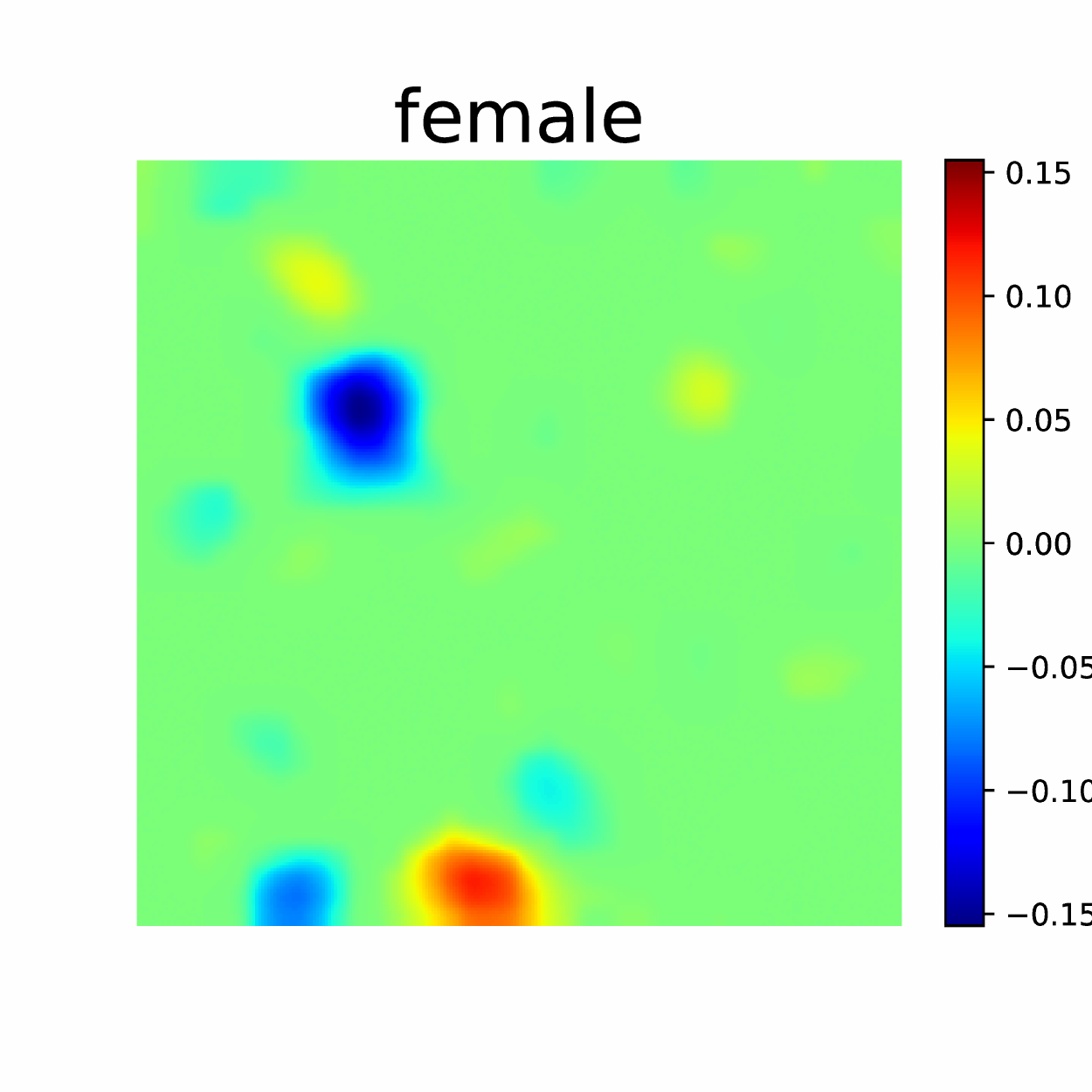} \\[-5pt]
\hskip-5pt
\includegraphics[width=.14\textwidth]{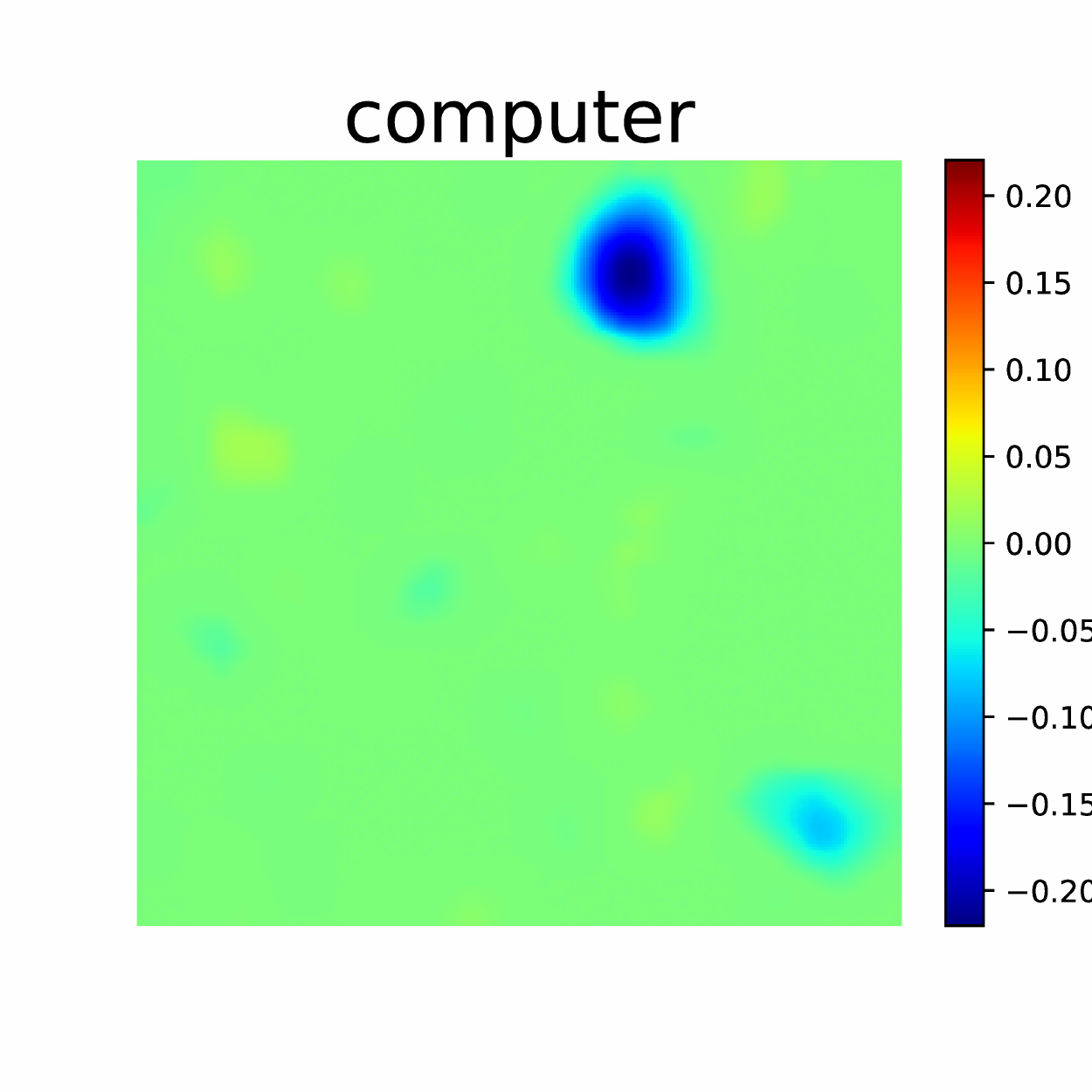} &
\includegraphics[width=.14\textwidth]{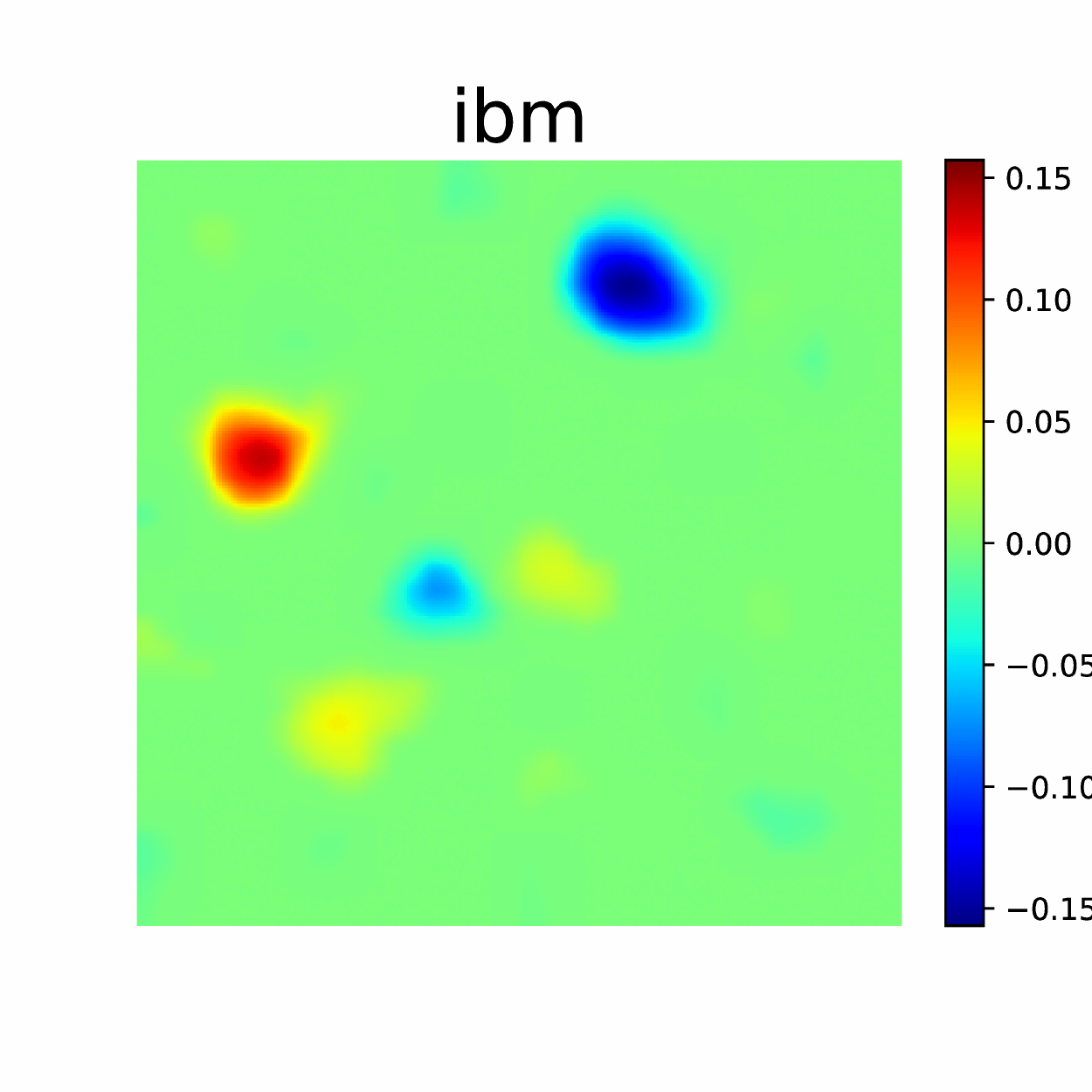} &
\includegraphics[width=.14\textwidth]{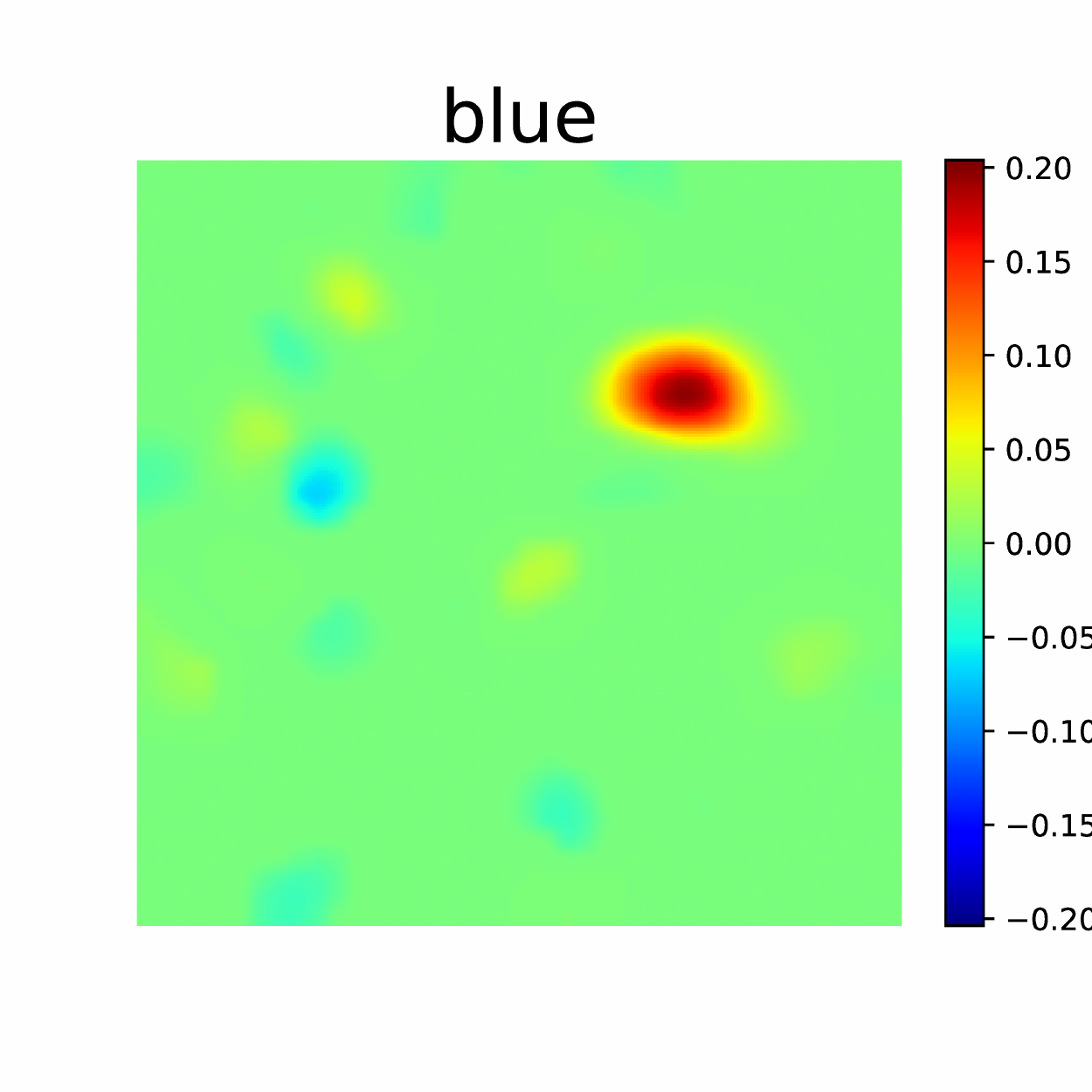} &
\includegraphics[width=.14\textwidth]{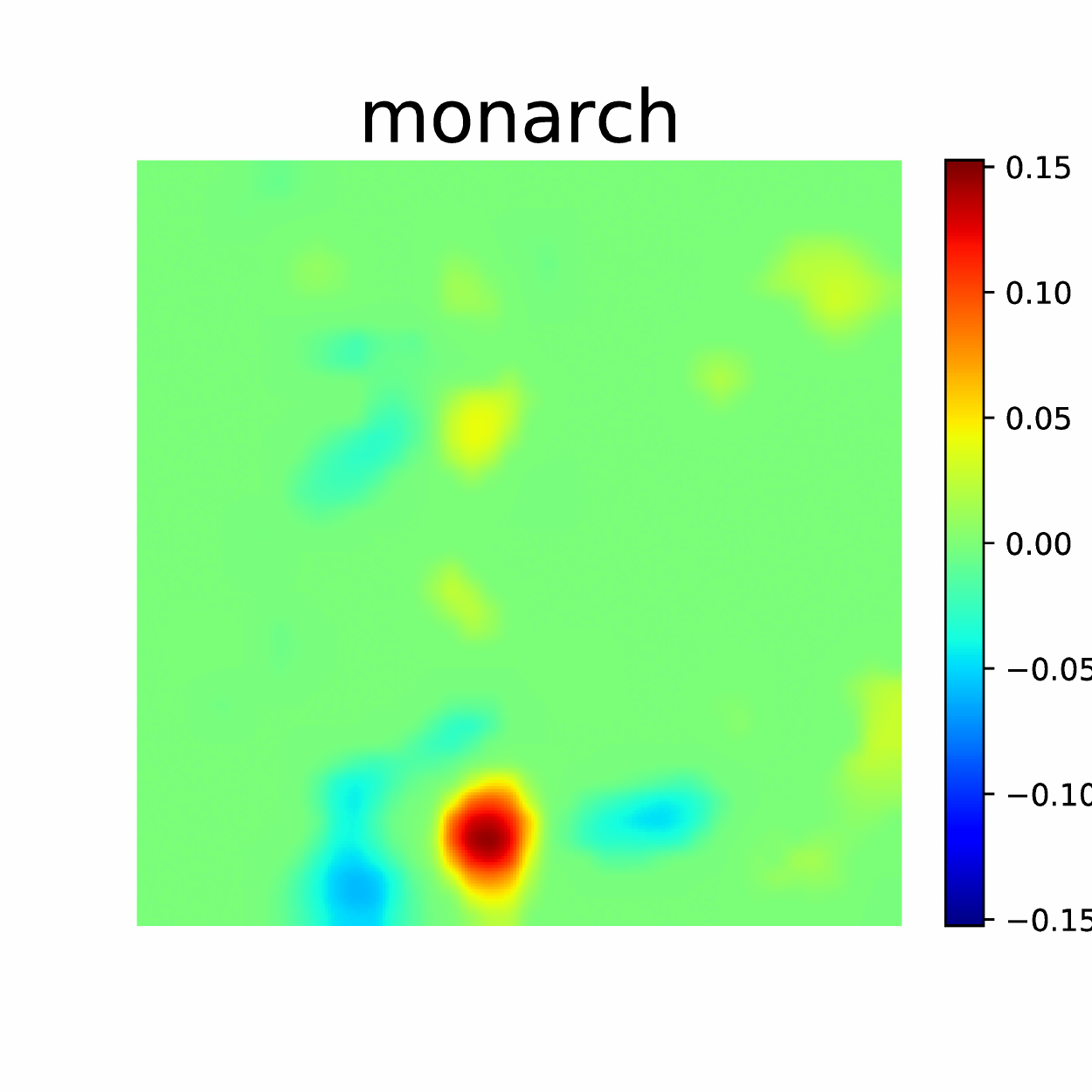} \\[-5pt]
\hskip-5pt
\includegraphics[width=.14\textwidth]{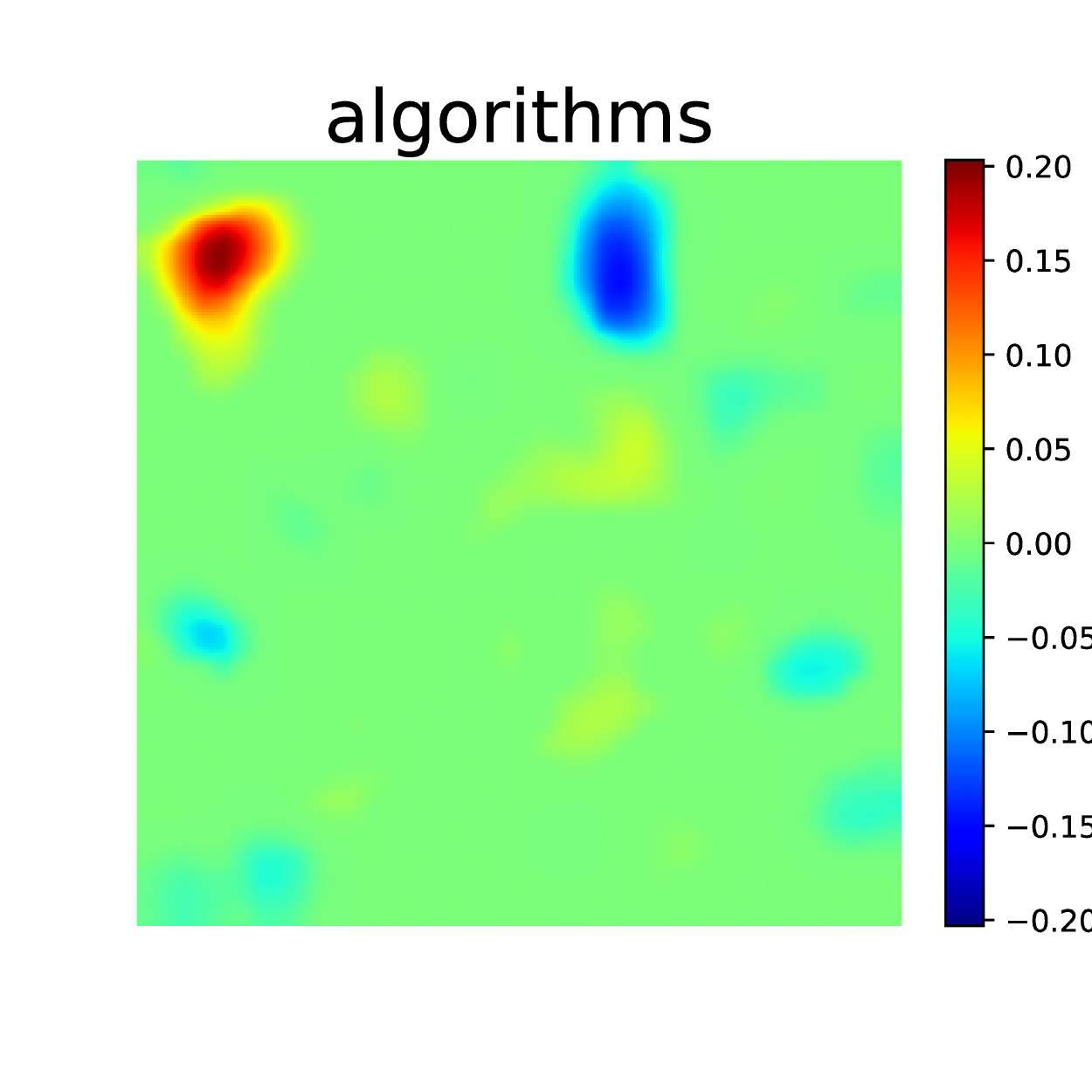} &
\includegraphics[width=.14\textwidth]{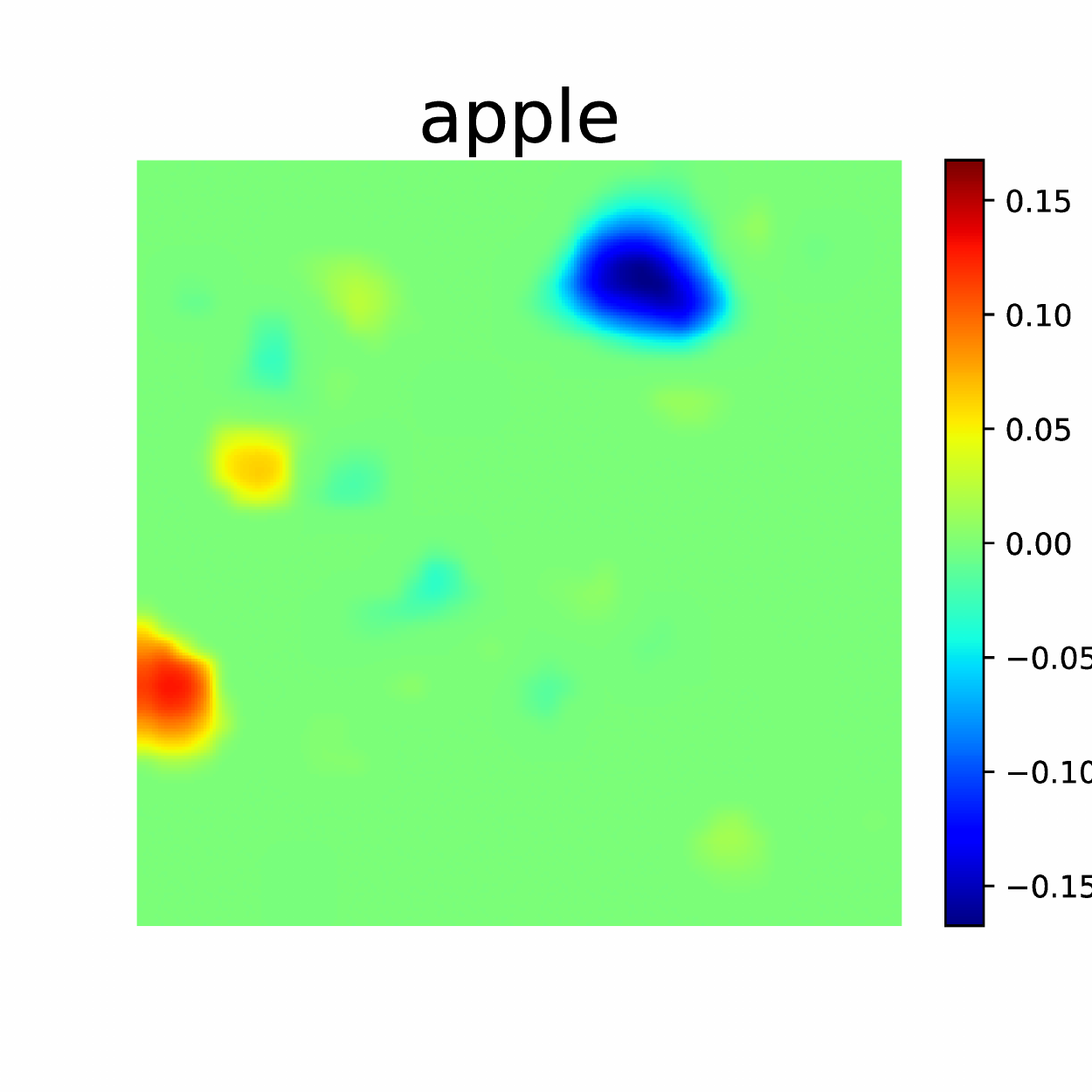} &
\includegraphics[width=.14\textwidth]{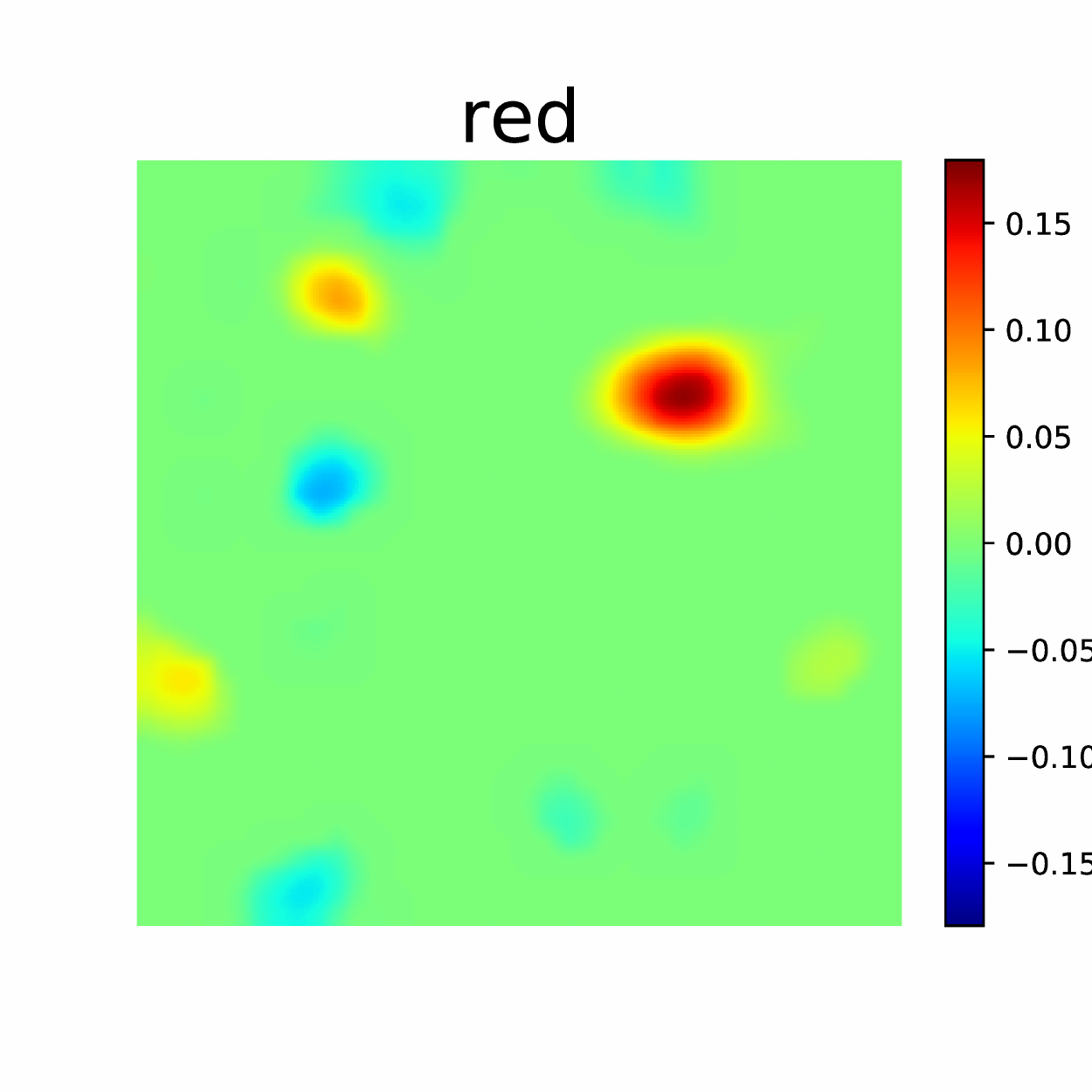} &
\includegraphics[width=.14\textwidth]{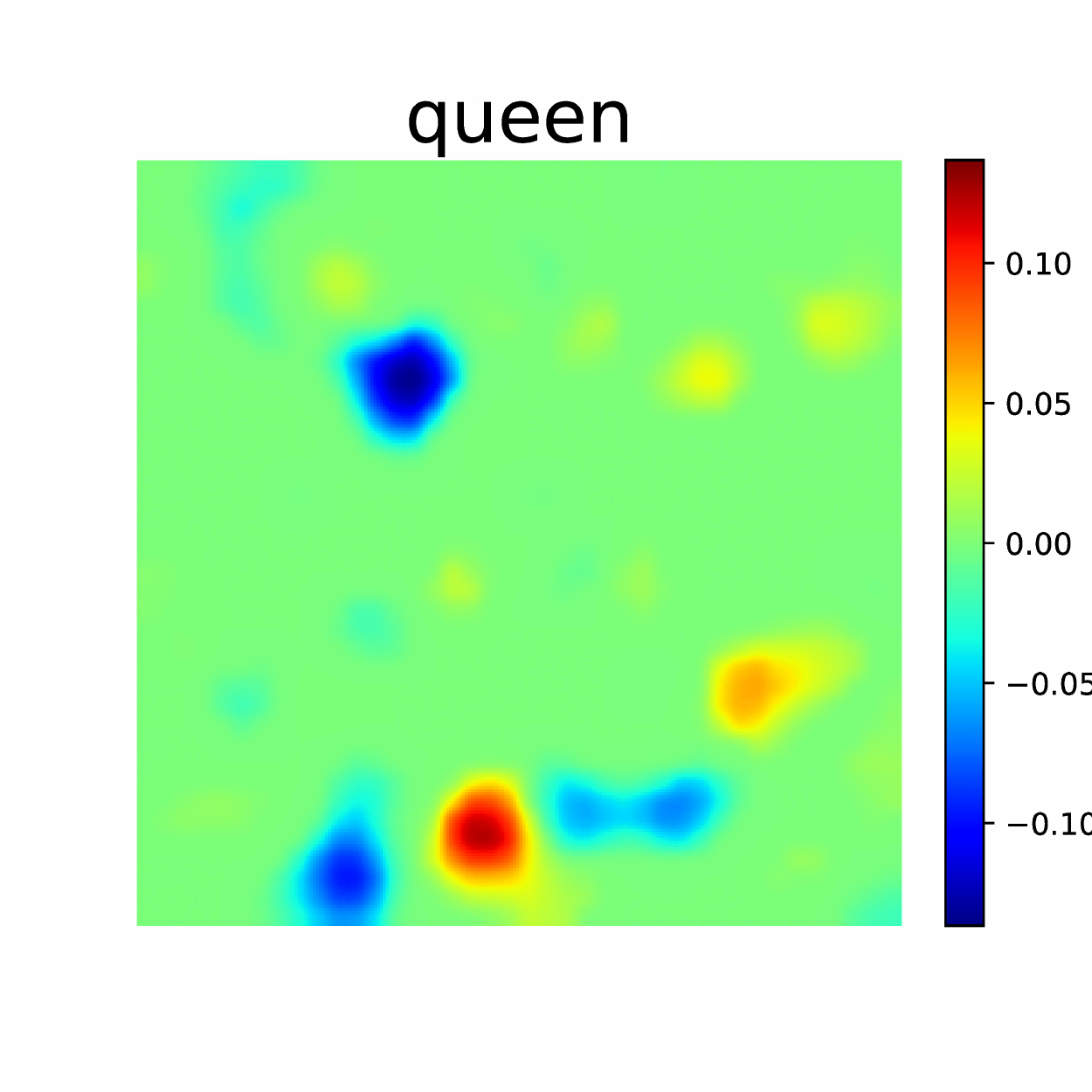}\\[-5pt]
\hskip-5pt
\includegraphics[width=.14\textwidth]{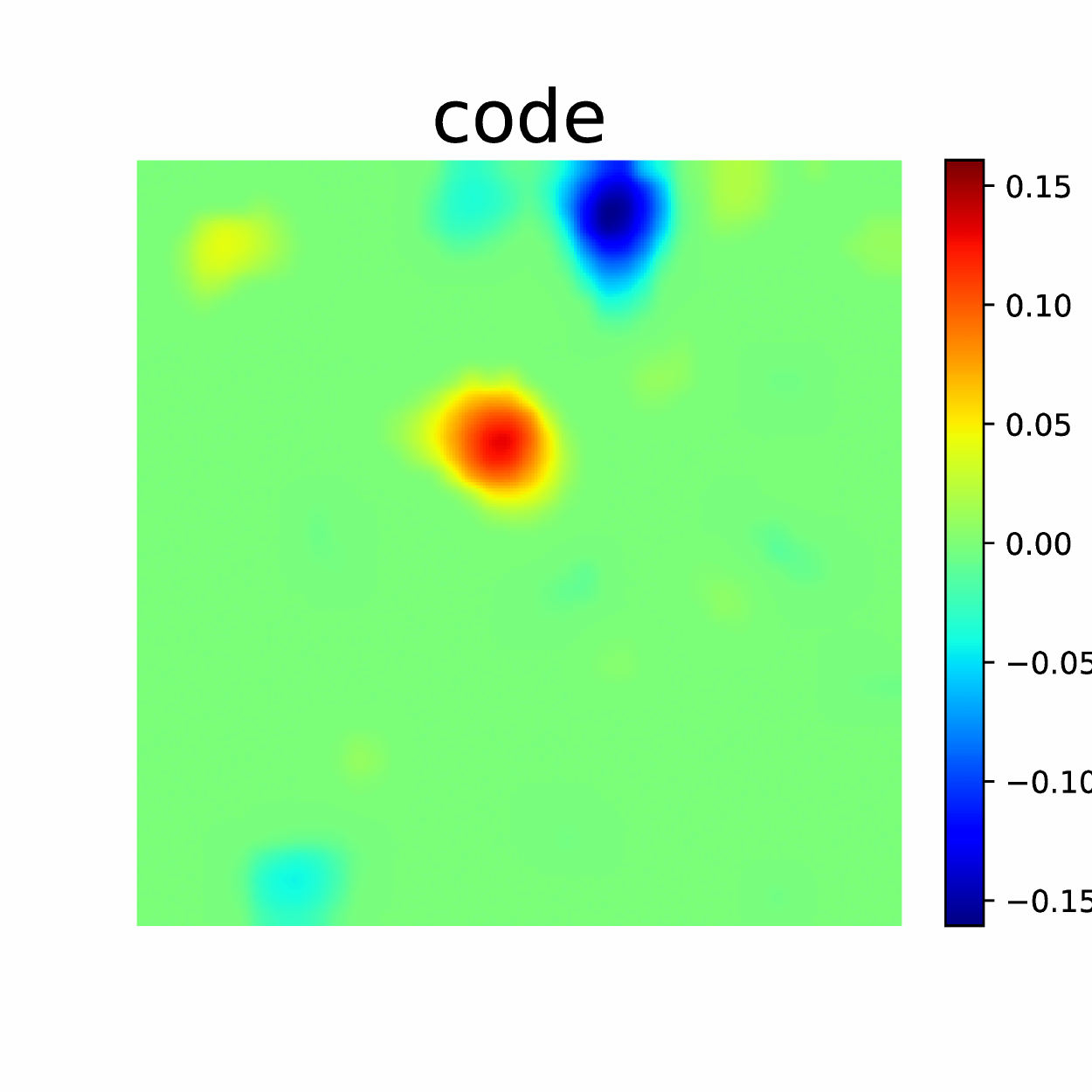} &
\includegraphics[width=.14\textwidth]{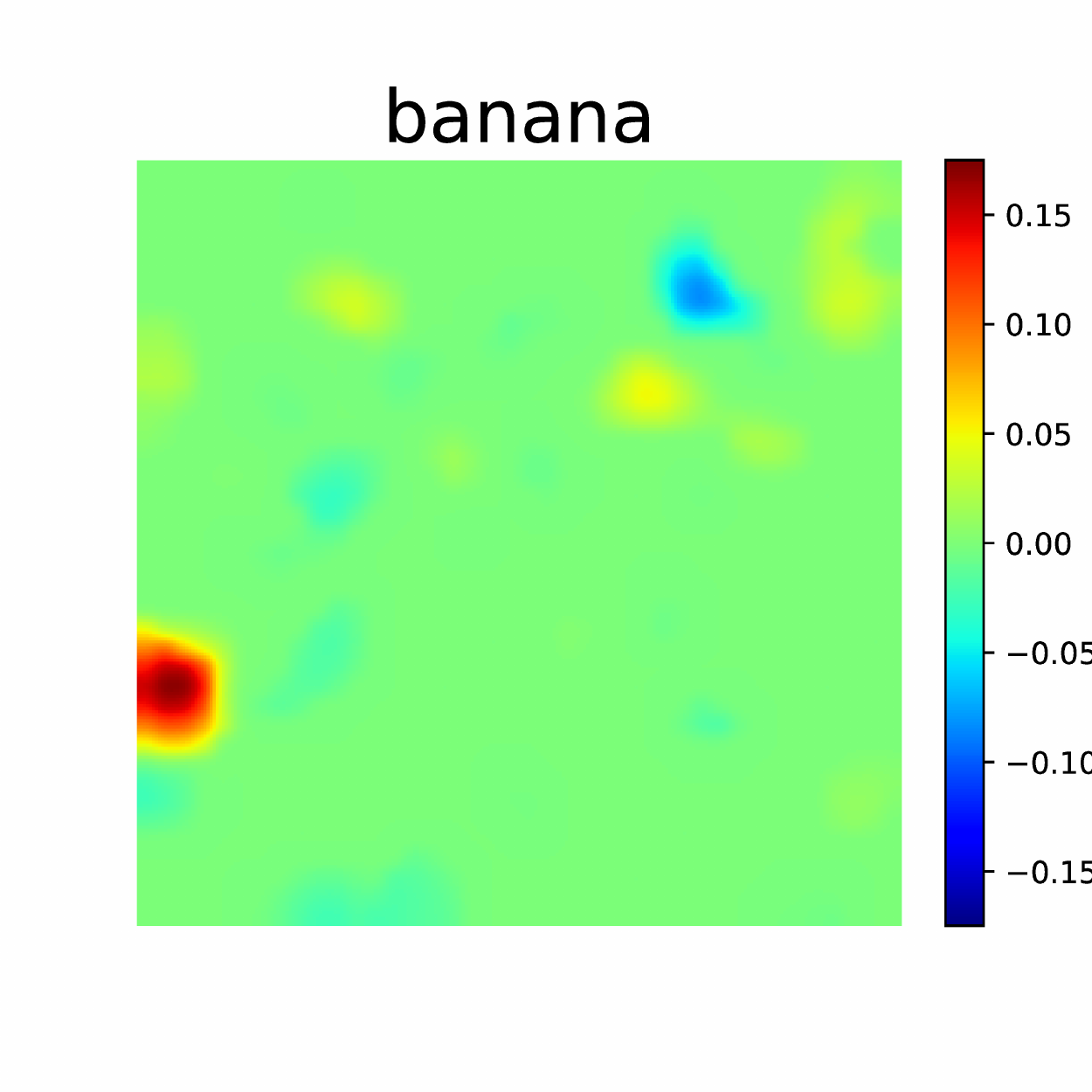} &
\includegraphics[width=.14\textwidth]{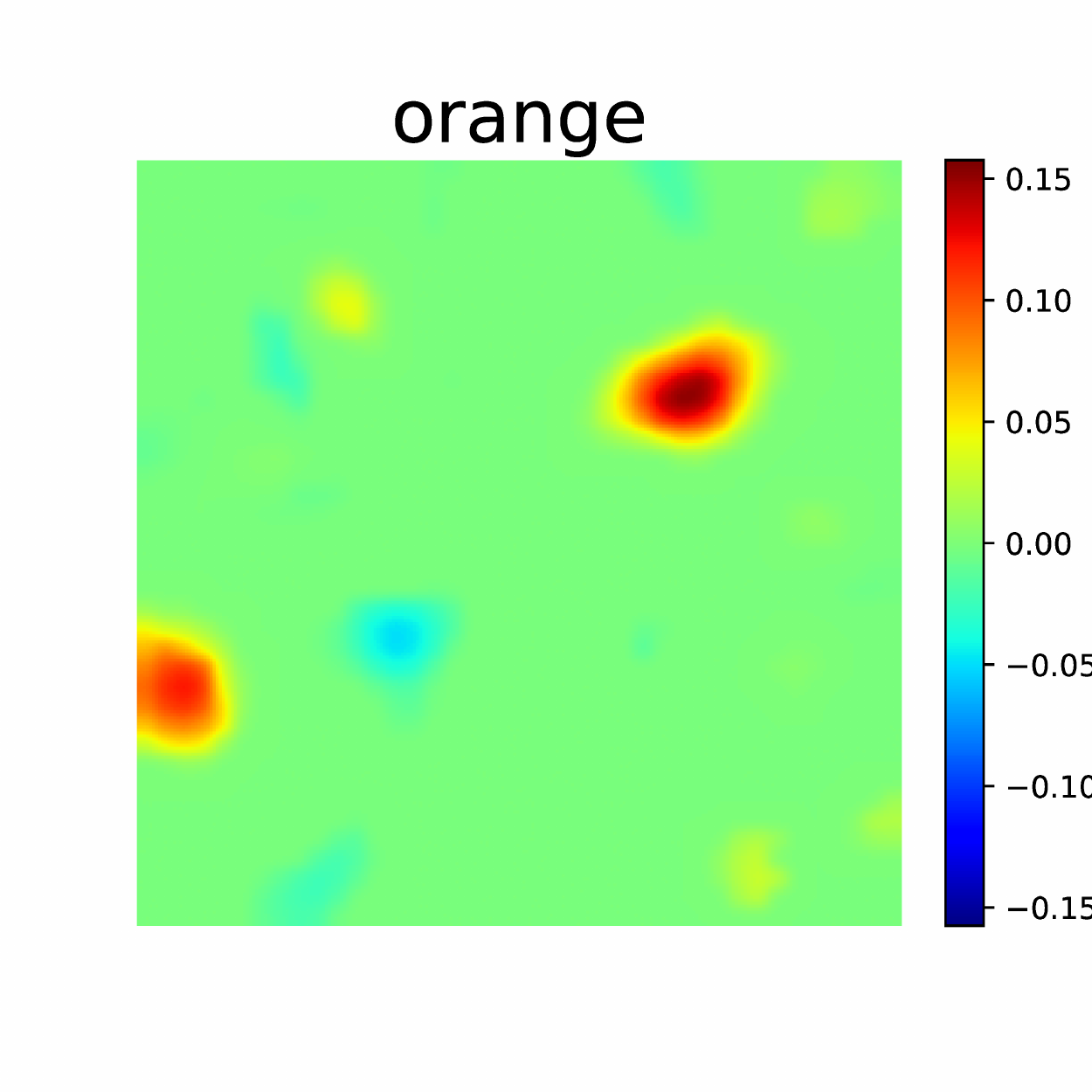} &
\includegraphics[width=.14\textwidth]{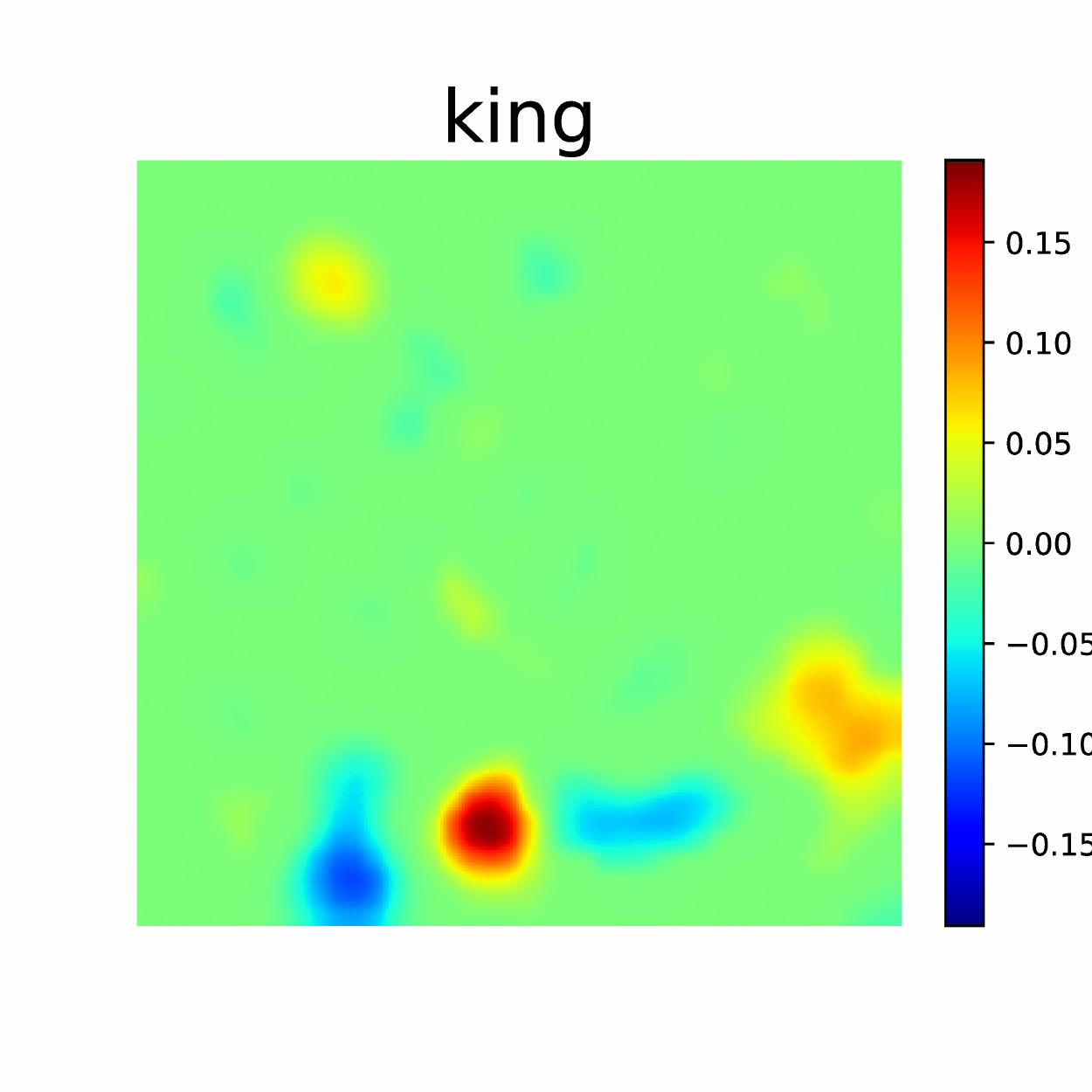}
\end{tabular}
\vskip1pt
\caption{Tunings of a grid of $40\times 40$ excitatory neurons from stimuli that are
embedding vectors of words, sampled from the unigram distribution
on a large corpus of text (Google 1-grams). Upper left: each neuron is mapped
to a color by taking the top 3-principal components of the receptive fields as an RGB value. Lower left: The activations are high dimensional sparse vectors, arranged in a 2-dimensional sheet; a standard 2-dimensional t-SNE visualization distorts the embedding distances. Right: Neural activations for selected groups of words.}
\label{fig:textcolortable}
\vskip-1pt
\end{figure}
}

The activation threshold $\lambda$
is set to achieve a target sparsity of $20\%$, meaning that $20\%$ of the neurons are active, on average,
after the iterative soft-thresholding algorithm has converged. The diagonal terms, corresponding to
voltage leak in the LIF model, are set to satisfy
\begin{equation}
d_{i,j} =
\sum_{(i',j') \in E(i,j)} W_{ij;i'j'}^E +  \sum_{(i',j') \in I(i,j)} W_I
\end{equation}
so that the optimization is convex; the algorithm \eqref{eq:ista} converges in under 15 steps.

Figure~\ref{fig:imagecodewords}
shows examples of the receptive fields that result.
While the algorithm is not directly minimizing the reconstruction error, the
error decreases and stabilizes. The tunings
are organized into regions resembling ``pinwheel'' patterns with respect to the orientations of the edge detectors
that have been measured experimentally \citep{crair}.
Figure~\ref{fig:imagecodewords} shows four image patches in a ``saccade'' along the edge of a piece of driftwood, and the neural responses.



\section{Organization of tunings to natural language text}
\label{sec:text}

In this section, we describe the results of applying the same algorithm and network configuration to natural language text. For our experiments with images, the stimulus is a random image patch from natural images. In this new
setting, the stimulus is a vector representation of a word sampled from the unigram distribution of a large corpus of natural language text. We consider two types of vector representations, which are closely related. In the first,
the vector is the 100-dimensional word embedding vector obtained by running the GloVe algorithm \citep{pennington2014glove}. We limit the vocabulary to the top 55,529 words in frequency, which can be expected to reduce but not eliminate the well known societal biases that are present in large-scale word embeddings \cite{kalai}. Sparse coding has previously been applied to word embeddings \citep{yogatama,faruqui,templeton}; in this work we are focused on the topographic map of semantic meaning that emerges from the organization of the receptive fields. We also ran experiments where the vector is the first layer in a LSTM recurrent neural network trained to predict the sequence of words
in a large corpus of text \citep{hochreiter1997long,ott2019fairseq}; the results are similar and discussed further in the supplement.
The structure of the excitatory and inhibitory network of neurons responding to these stimuli is the same as used above---a grid of $40\times 40$ excitatory neurons is paired with a grid of $40\times 40$ inhibitory neurons, with a neuron at a given grid point being
connected to all excitatory neurons within a radius $d_E=3$ and to all inhibitory neurons with a radius $d_I = 5$. The algorithm is run for 50,000 gradient steps.

Figure~\ref{fig:textcolortable} illustrates the tunings in the grid of $40\times 40$ excitatory neurons from stimuli as described above using the GloVe embeddings; the organization of receptive fields when the stimuli are embedding vectors from training an LSTM language model is qualitatively similar.
To visualize the map, principal components analysis is carried out for the collection
of 1,600 receptive fields. Each neuron is mapped
to a color by taking the top 3-principal components as an RGB value.  The map
shows local regions where the color gradient changes slowly,
implying a gradual change in the semantic meaning represented by the underlying neurons.
The groups of similarly colored neurons correspond to edge detectors with similar
spatial orientations and scales in the image model.

\afterpage{
\begin{figure*}[ht]
  \begin{center}
    \begin{tabular}{ccc}
    \raisebox{20pt}{\includegraphics[width=.3\textwidth]{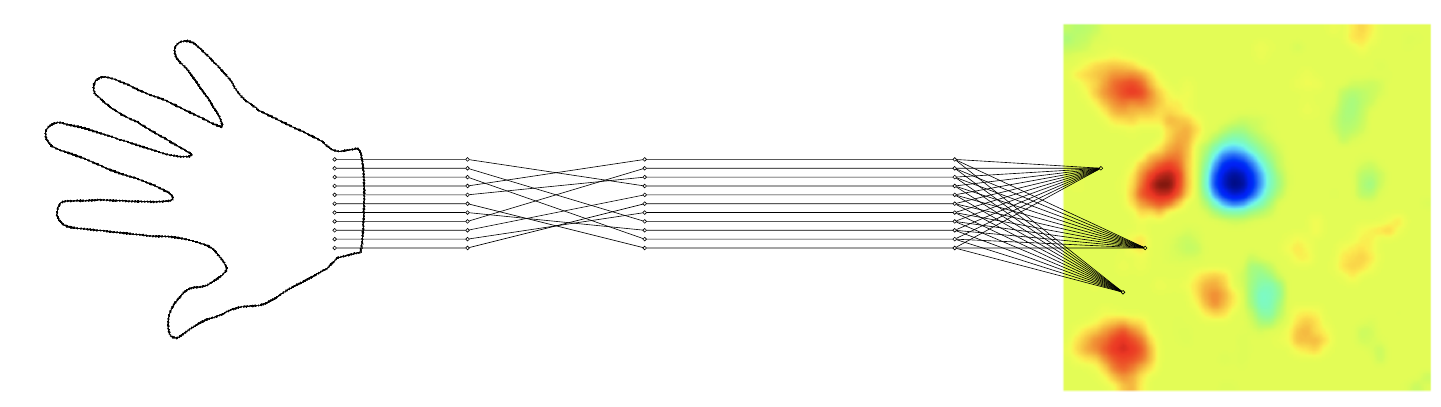}}&
    \includegraphics[width=.3\textwidth]{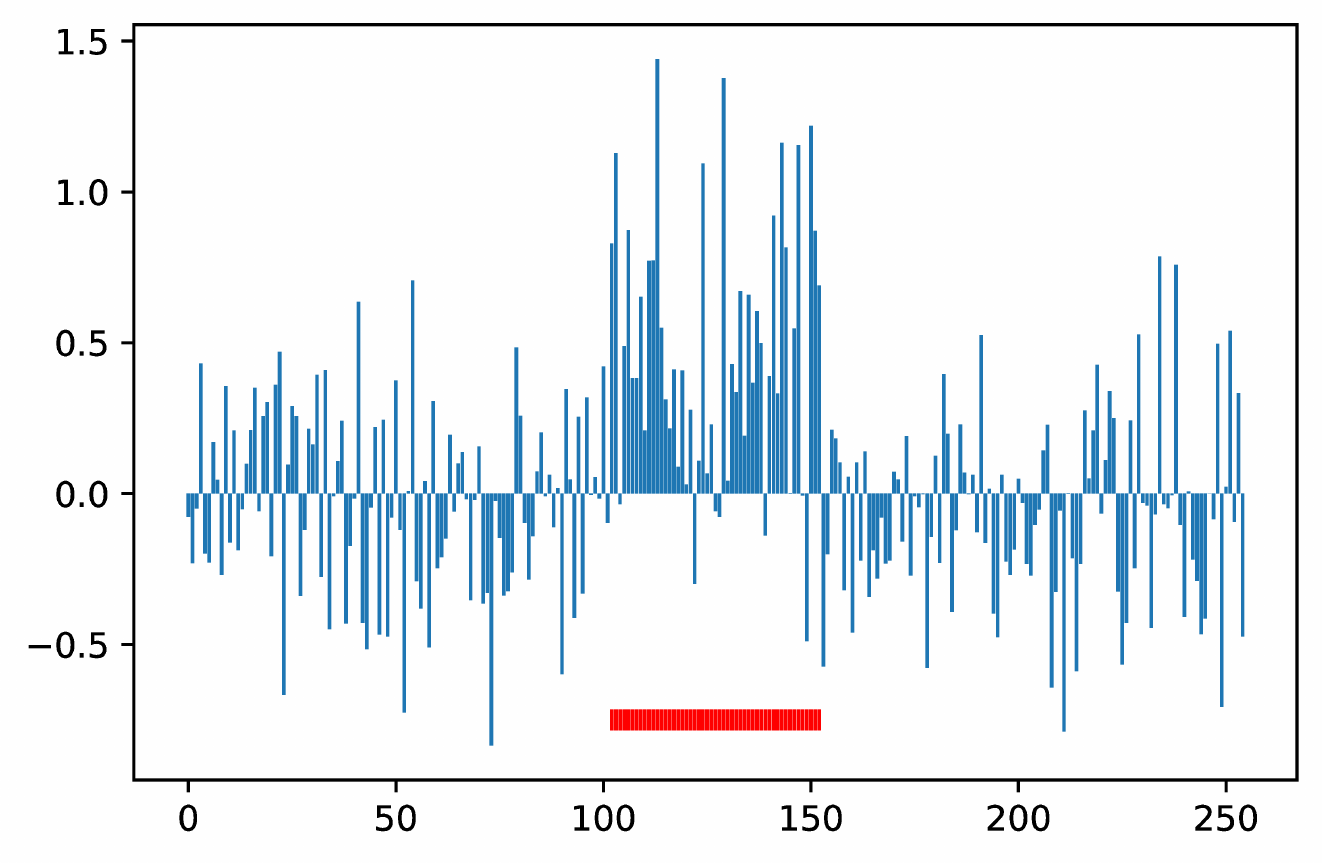} &
    \includegraphics[width=.3\textwidth]{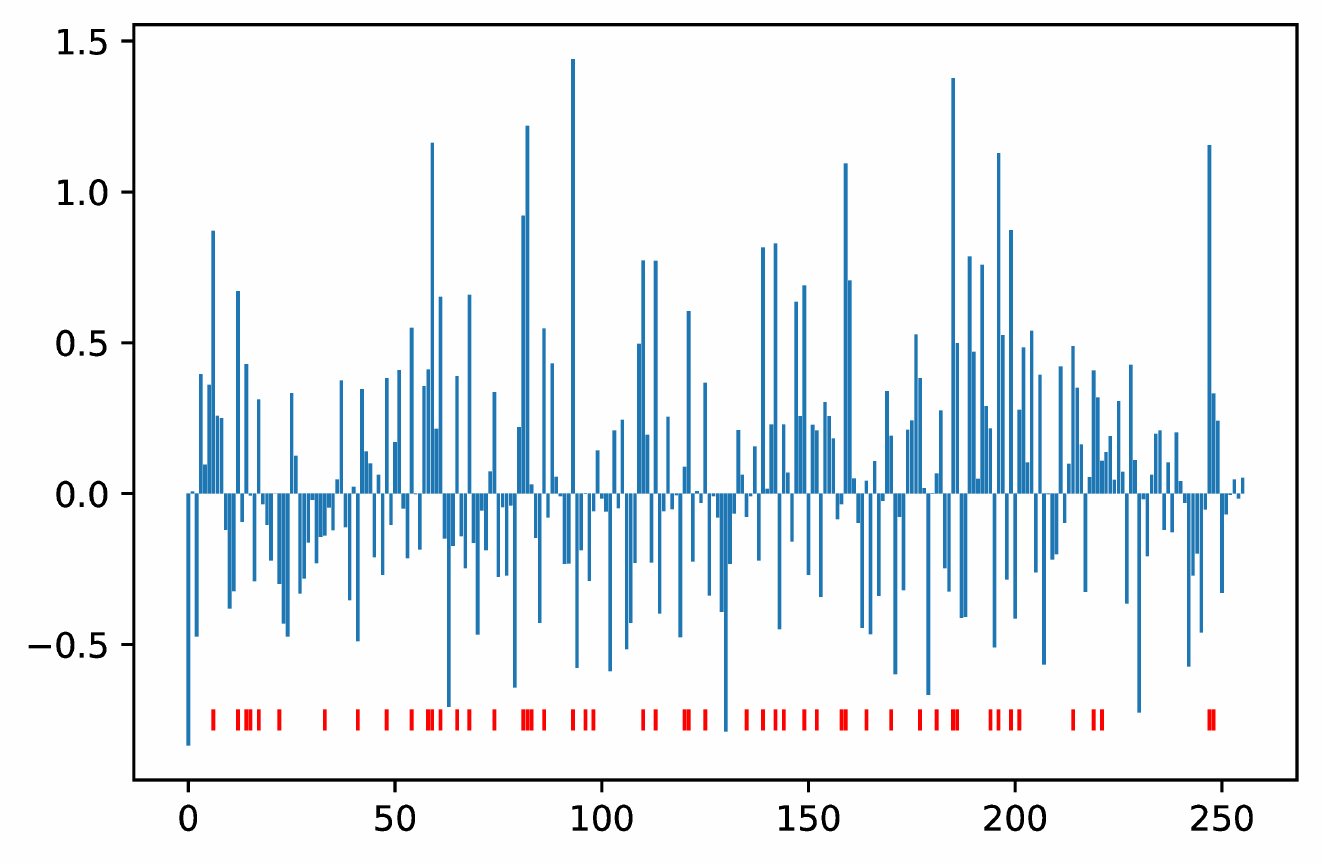} \\
    \end{tabular}
  \end{center}
  \caption{The nerves in each of the five ``fingers''
  generate random Gaussian outputs in the baseline state;
  when a finger is touched, uniform noise
  is added to the signal. The nerves across the fingers are
  rearranged according to a fixed, random permutation,
  resulting in a 256-dimensional signal received at each excitatory neuron.
  Center: A signal generated when digit 3 is touched; Right: the stimulus received at the excitatory neurons.
  \label{fig:cartoon}
  }
\vskip-10pt
\renewcommand{\arraystretch}{3}
\begin{center}
\vskip-10pt
\begin{tabular}{cc}
\begin{tabular}{c}
  \includegraphics[width=.18\textwidth]{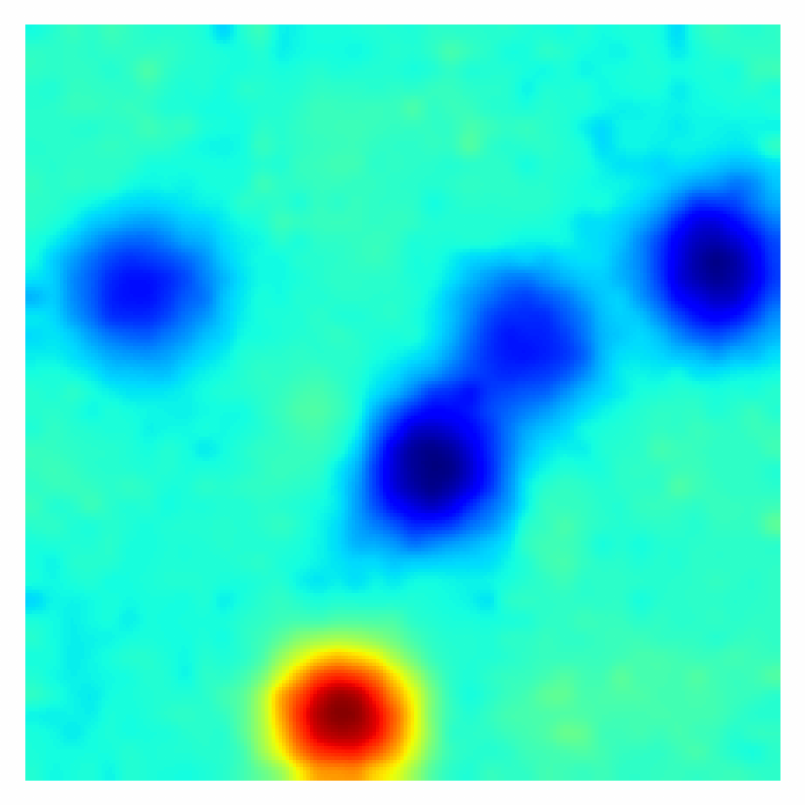}\\
  \includegraphics[width=.18\textwidth]{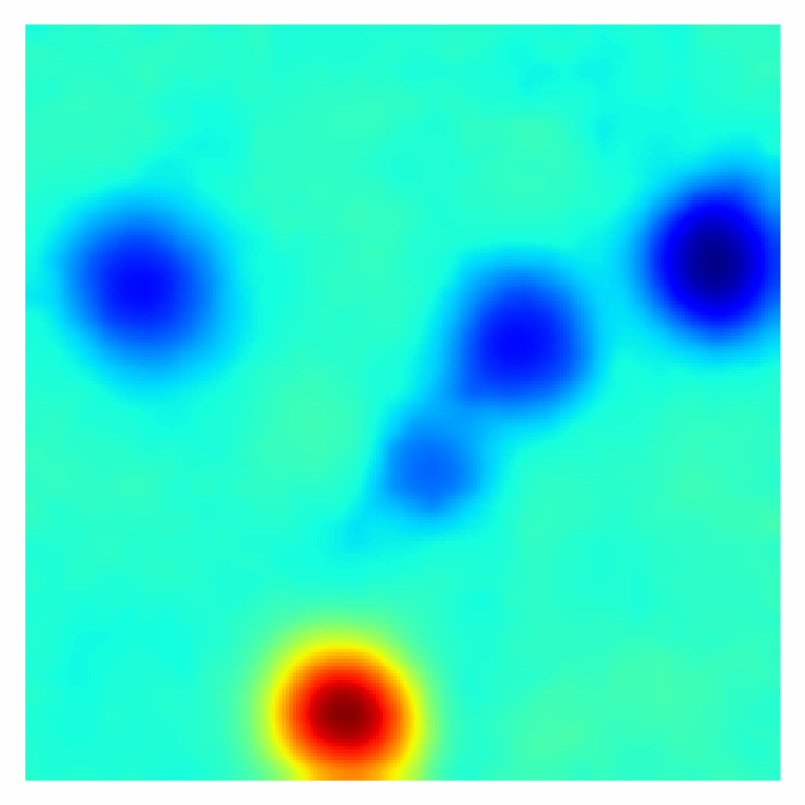} \\[-20pt]
\end{tabular}
&
\begin{tabular}{cccc}
{\small 1} & \small 2 & \small 3 & \small{1 \& 3}\\
\hskip-5pt
\includegraphics[width=.14\textwidth]{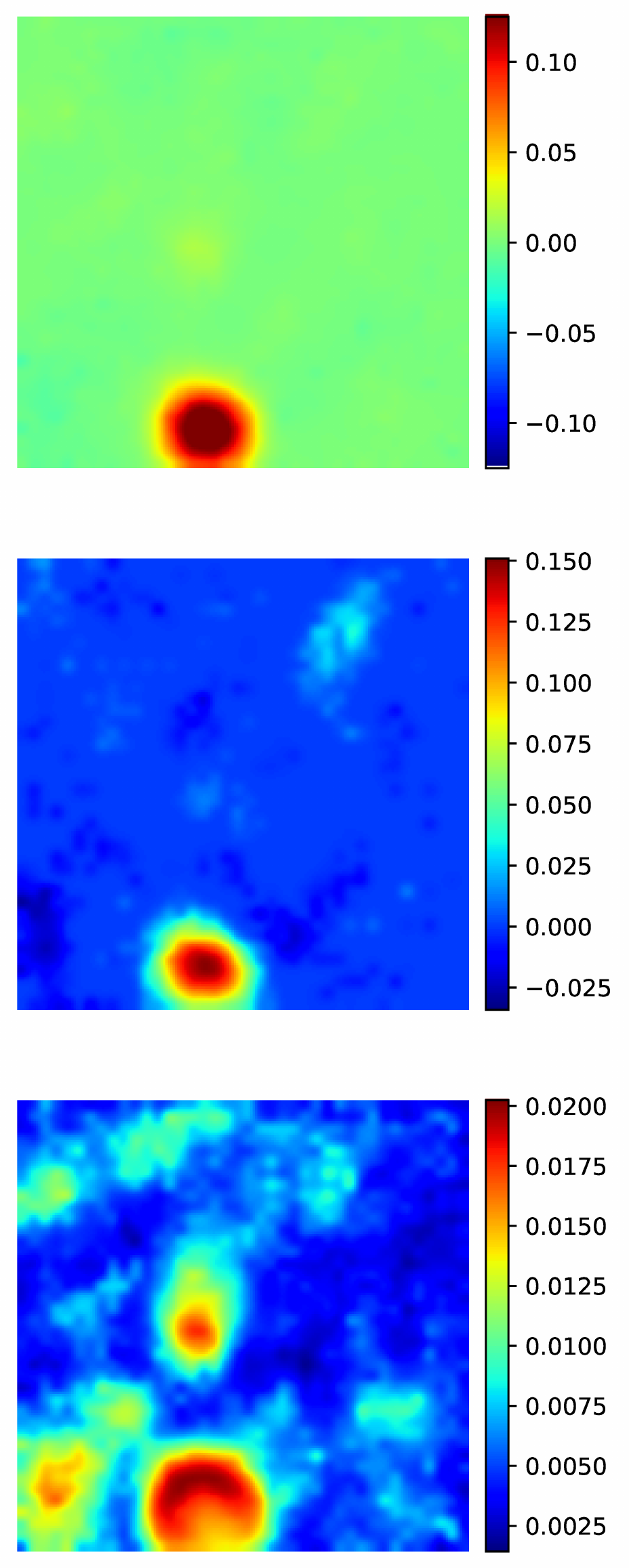} &
\includegraphics[width=.14\textwidth]{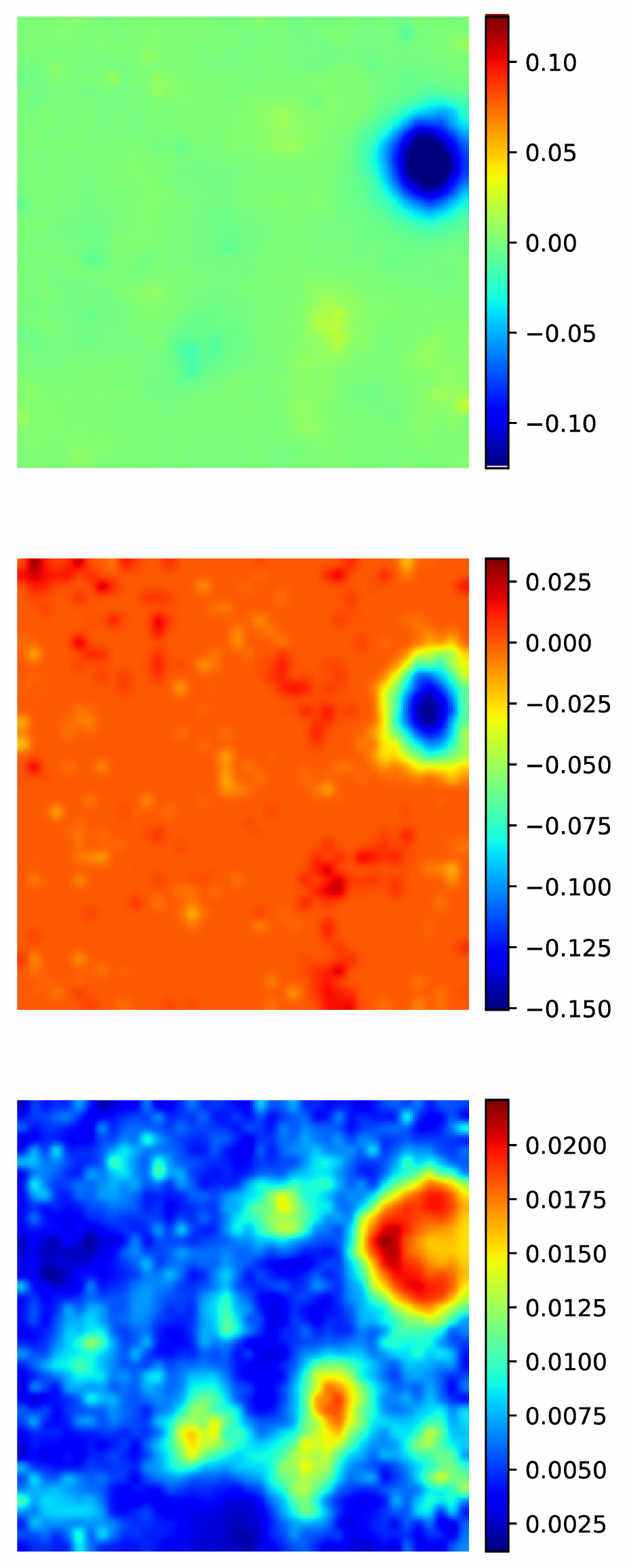} &
\includegraphics[width=.14\textwidth]{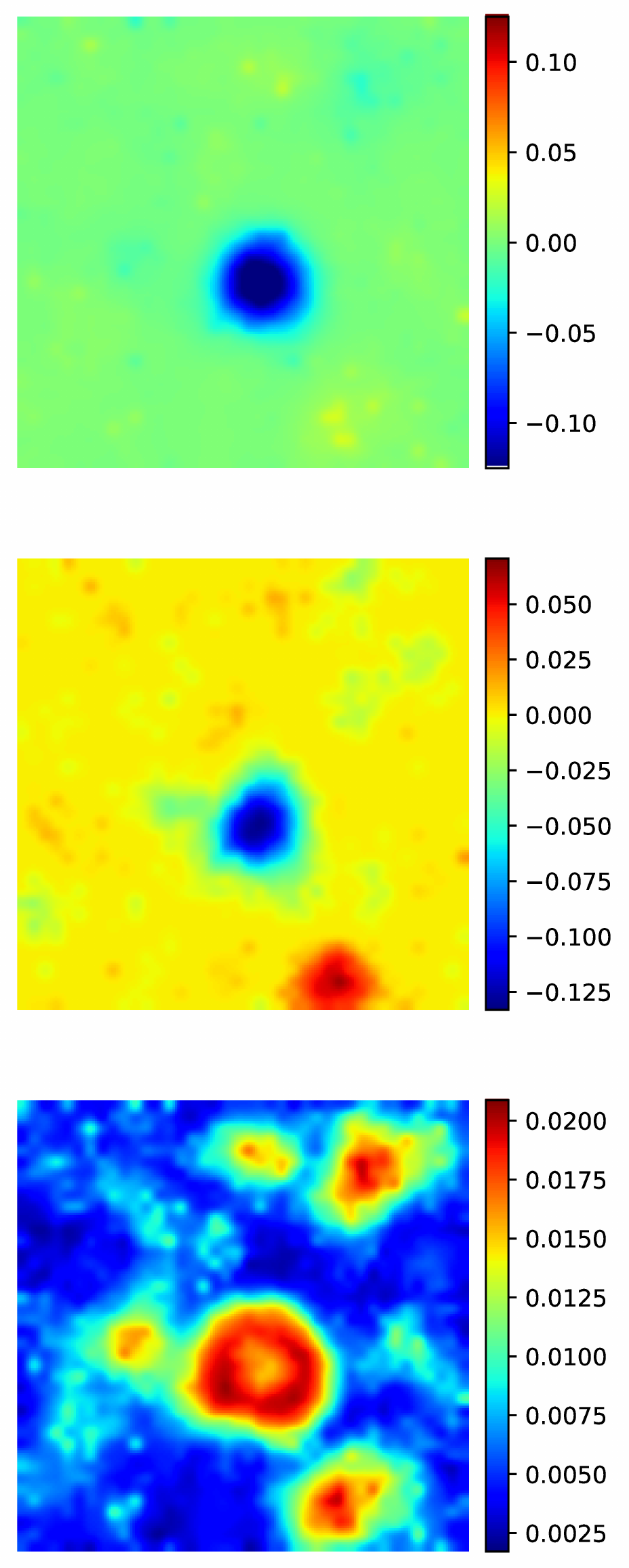} &
\includegraphics[width=.14\textwidth]{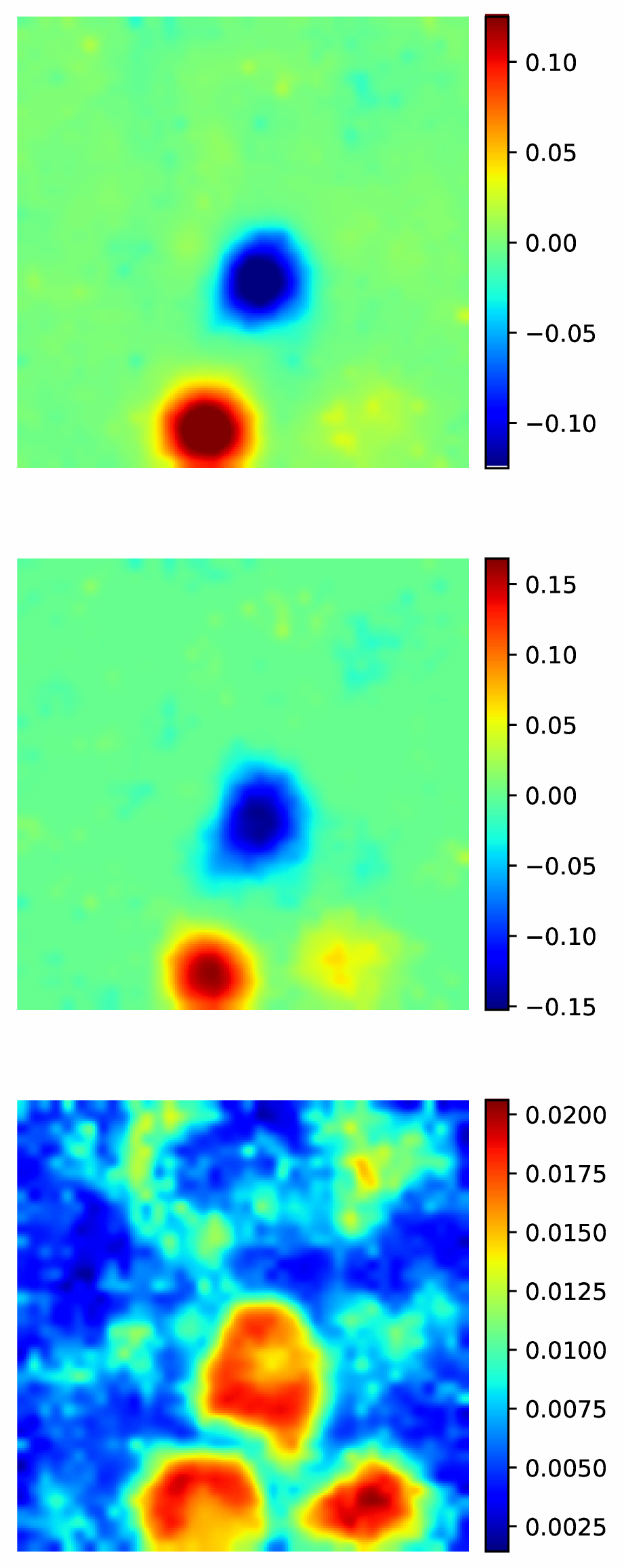}
\end{tabular}
\end{tabular}
\end{center}
\caption{Neuronal tunings to synthetic touch data. Upper left:
Average of all activations, showing localized receptive fields
for each digit.
Right plots: Average activation (top row), activation for random stimulus (middle row) and standard deviation of activations (bottom row),
when a stimulus is a touch to digit 1, 2, or 3,
or to digits 1 and 3 simultaneously (last column).
Lower left: Map after digit 3 is no longer touched, but the
other digits are, showing diminished receptive field for that digit.}
\label{fig:fingermaps}
\vskip1pt
\end{figure*}
}

Figure~\ref{fig:textcolortable} also illustrates how the organization of the neuronal
tunings encodes these semantic differences. The activations for sets of related words are shown in columns. In
the first column, for example, ``algorithms'' are semantically ``equations for computers.'' Several other semantic relations can be seen to be at play by inspecting the different groups of words and their activations. Such relations
can be discovered by searching for words that have large components in specific semantic components.
While direct neural recordings are not available for language regions of the human brain, fMRI studies suggest a distributed semantic map \citep{huth}; in future work this
might be compared with the emergent structure in our models.


\section{A model of sensory perception}
\label{sec:fingers}

In our last computational experiment, we investigate a simple model
of somatosensory perception in a toy computational model of a hand with five fingers.
According to the model, in the default state the nerves in each of the five ``fingers''
generate random Gaussian outputs; when a finger is touched, uniformly distributed noise
is added to the signal. The nerves across the fingers are
rearranged according to a fixed, random permutation,
resulting in a 256-dimensional signal received at the excitatory neurons.
Seven possible input types are generated; either a single digit 1--5
is ``touched'' with equal probability, or digits 1 and 2 or 1 and 3 are
simultaneously touched.
Figure~\ref{fig:cartoon} shows an example where digit 3 is touched.

Similar to the image and text experiments, each excitatory neuron in a $40\times 40$
grid receives these 256-dimensional stimuli. In this case, the network parameters
are such that excitatory neurons are connected to each neuron within a radius of $d_E = 10$,  and inhibitory neurons are connected to each neuron within a radius of $d_I = 13$. The resulting
receptive fields are shown in Figure~\ref{fig:fingermaps}. The neurons that
respond to the touch of each of the five digits are clearly localized in a coherent region,
each being of roughly equal size. Note that this is an unsupervised learning
algorithm. However, given the neuronal map, a supervised algorithm would
learn to name the finger touches with just a handful of training examples.
We find that when the same recurrent geometry is used in an artificial neural network in a purely supervised fashion, to predict which digit is touched, the
topographic map is not well formed.

Next, the weights for the 1,600 neurons continue to be tuned, but we remove inputs corresponding to touches of digit 3 individually and digits 1 and 3 together,
as if digit 3 were to be disabled. As seen in Figure~\ref{fig:fingermaps}, the receptive fields for digits 1, 2, 4, and 5 remain stable. However, the size of the region that codes for signals from digit~3 begins to shrink, in a type of ``use it or lose it'' phenomenon, as those neurons are no longer activated. Notably, for this setting of network parameters the other digits do not recruit the neurons previously coding for digit~3.
In this way, such activation models might serve as a computational tool
to complement studies on reorganization of neural maps in the primary somatosensory cortex
\citep{makin, wesselink}.

\section{Discussion}
\label{sec:discuss}

This paper studied how the local patterns of excitation and inhibition that can generate spontaneous neural waves may also provide a computational mechanism for the emergent organization of receptive fields. Starting from a leaky integrate-and-fire model for modeling neural waves, we developed a new model where the firing of neurons is abstracted into a continuous level of activation. We formulated the neural dynamics of this model as an iterative procedure for solving a convex optimization, coupled with an algorithm for carrying out stochastic gradient descent over the feedforward weights. We found that the emergent receptive fields resemble known pinwheel patterns in V1 when trained on natural images, exhibit a type of compositional semantics when trained on word embeddings for large language models, and form a topographic map of the hand when trained on a simplified model of somatosensory input. For the image and text experiments, the entire training algorithm takes roughly 10 hours on a standard CPU machine. Since the Laplacian is based on a regular graph and can be expressed as a convolution, the computation could be accelerated with GPUs. 

The neural activations in our model are driven by the local, lateral
excitations and inhibitions, while the gradient update uses reconstruction error. We have observed empirically that the overall algorithm iteratively decreases the error. It is important to develop a mathematical analysis of this phenomenon in future work.
Recent work has shown how convergence can be obtained in other contexts for two-layer networks that do not directly minimize a loss function
\citep{song}. Another interesting direction to explore is
``pretraining'' with internal stimuli from retinal waves, to shape
the V1 receptive fields before they are tuned on natural, external stimuli.

While related work in machine learning has considered models of
complex cells and structured inhibition \citep{hoyer,gregor2010emergence}
in the context of sparse coding of natural images, the framework developed here can be seen as more basic
from a biological perspective. The sensory model explored here is limited; it would be of interest to build on these results with more realistic models. For text data, our experiments could in future work be compared with
clustering-based models derived from fMRI studies of language perception,
which have found that semantic representation of language is widely distributed
\citep{huth}. Finally, we note that while algorithms like t-SNE are commonly used to
 visualize vector representations of objects \citep{vanDerMaaten2008}, the algorithms developed here provide an alternative way of interpreting and visualizing any such representation, separate from any motivation from neuroscience.


\section*{Acknowledgements}

We thank Damon Clark, Michael Crair, Samuel McDougle, and Baohua Zhou for helpful comments on this work.
Research supported in part by NSF grant CCF-1839308 and NSF grant DMS-2015397.

\setlength{\bibsep}{8pt plus 0.3ex}
\bibliographystyle{apalike}
\bibliography{neural-waves,camri,feedback_alignment,text}

\begin{thebibliography}{}

\bibitem[Ackman et~al., 2012]{ackman2012retinal}
Ackman, J.~B., Burbridge, T.~J., and Crair, M.~C. (2012).
\newblock Retinal waves coordinate patterned activity throughout the developing
  visual system.
\newblock {\em Nature}, 490(7419):219--225.

\bibitem[Ackman and Crair, 2014]{ackman2014role}
Ackman, J.~B. and Crair, M.~C. (2014).
\newblock Role of emergent neural activity in visual map development.
\newblock {\em Current opinion in neurobiology}, 24:166--175.

\bibitem[Akrout et~al., 2019]{akrout}
Akrout, M., Wilson, C., Humphreys, P., Lillicrap, T., and Tweed, D.~B. (2019).
\newblock Deep learning without weight transport.
\newblock In Wallach, H., Larochelle, H., Beygelzimer, A., d'Alch\'{e} Buc, F.,
  Fox, E., and Garnett, R., editors, {\em Advances in Neural Information
  Processing Systems}, volume~32. Curran Associates, Inc.

\bibitem[Beck and Teboulle, 2008]{beck:ista}
Beck, A. and Teboulle, M. (2008).
\newblock A fast iterative shrinkage-thresholding algorithm for linear inverse
  problems.
\newblock {\em SIAM J. Imaging Sciences}, 2(1):183--202.

\bibitem[Bell and Sejnowski, 1997]{Bell:97}
Bell, A.~J. and Sejnowski, T.~J. (1997).
\newblock The ``independent components'' of natural scenes are edge filters.
\newblock {\em Vision Research}, 37(23):3327--3338.

\bibitem[Bellec et~al., 2019]{bellec}
Bellec, G., Scherr, F., Hajek, E., Salaj, D., Legenstein, R., and Maass, W.
  (2019).
\newblock Biologically inspired alternatives to backpropagation through time
  for learning in recurrent neural nets.

\bibitem[Benucci et~al., 2007]{benucci2007standing}
Benucci, A., Frazor, R.~A., and Carandini, M. (2007).
\newblock Standing waves and traveling waves distinguish two circuits in visual
  cortex.
\newblock {\em Neuron}, 55(1):103--117.

\bibitem[Bolukbasi et~al., 2016]{kalai}
Bolukbasi, T., Chang, K.-W., Zou, J., Saligrama, V., and Kalai, A. (2016).
\newblock Man is to computer programmer as woman is to homemaker? {D}ebiasing
  word embeddings.
\newblock In {\em Advances in Neural Information Processing Systems}, pages
  4349--4357.

\bibitem[Bonhoeffer and Grinvald, 1991]{bonhoeffer:1991}
Bonhoeffer, T. and Grinvald, A. (1991).
\newblock Iso-orientation domains in cat visual cortex are arranged in
  pinwheel-like patterns.
\newblock {\em Nature}, 353(6343):429--431.

\bibitem[Bonhoeffer and Grinvaldi, 1993]{bonhoeffer:1993}
Bonhoeffer, T. and Grinvaldi, A. (1993).
\newblock The layout of iso-orientation domains in area 18 of cat visual
  cortex: {O}ptical imaging reveals a pinwheel-like organization.
\newblock {\em The Journal of Neuroscience}, 13(10):4157--4180.

\bibitem[Bressloff, 2000]{bressloff2000traveling}
Bressloff, P.~C. (2000).
\newblock Traveling waves and pulses in a one-dimensional network of excitable
  integrate-and-fire neurons.
\newblock {\em Journal of Mathematical Biology}, 40(2):169--198.

\bibitem[Brunel, 2000]{brunel2000dynamics}
Brunel, N. (2000).
\newblock Dynamics of sparsely connected networks of excitatory and inhibitory
  spiking neurons.
\newblock {\em Journal of computational neuroscience}, 8(3):183--208.

\bibitem[Brunel and Wang, 2003]{brunel2003determines}
Brunel, N. and Wang, X.-J. (2003).
\newblock What determines the frequency of fast network oscillations with
  irregular neural discharges? {I. S}ynaptic dynamics and excitation-inhibition
  balance.
\newblock {\em Journal of neurophysiology}, 90(1):415--430.

\bibitem[Chemla et~al., 2019]{chemla2019suppressive}
Chemla, S., Reynaud, A., Di~Volo, M., Zerlaut, Y., Perrinet, L., Destexhe, A.,
  and Chavane, F. (2019).
\newblock Suppressive traveling waves shape representations of illusory motion
  in primary visual cortex of awake primate.
\newblock {\em Journal of Neuroscience}, 39(22):4282--4298.

\bibitem[Churchland et~al., 2010]{churchland2010stimulus}
Churchland, M.~M., Byron, M.~Y., Cunningham, J.~P., Sugrue, L.~P., Cohen,
  M.~R., Corrado, G.~S., Newsome, W.~T., Clark, A.~M., Hosseini, P., Scott,
  B.~B., et~al. (2010).
\newblock Stimulus onset quenches neural variability: a widespread cortical
  phenomenon.
\newblock {\em Nature neuroscience}, 13(3):369--378.

\bibitem[Crair et~al., 1997]{crair}
Crair, M.~C., Ruthazer, E.~S., Gillespie, D.~C., and Stryker, M.~P. (1997).
\newblock Ocular dominance peaks at pinwheel center singularities of the
  orientation map in cat visual cortex.
\newblock {\em Journal of Neurophysiology}, 77(6):3381--3385.

\bibitem[Davis et~al., 2020]{davis2020spontaneous}
Davis, Z.~W., Muller, L., Martinez-Trujillo, J., Sejnowski, T., and Reynolds,
  J.~H. (2020).
\newblock Spontaneous travelling cortical waves gate perception in behaving
  primates.
\newblock {\em Nature}, 587(7834):432--436.

\bibitem[Ermentrout and Kleinfeld, 2001]{ermentrout2001traveling}
Ermentrout, G.~B. and Kleinfeld, D. (2001).
\newblock Traveling electrical waves in cortex: {I}nsights from phase dynamics
  and speculation on a computational role.
\newblock {\em Neuron}, 29(1):33--44.

\bibitem[Faruqui et~al., 2015]{faruqui}
Faruqui, M., Tsvetkov, Y., Yogatama, D., Dyer, C., and Smith, N.~A. (2015).
\newblock Sparse overcomplete word vector representations.
\newblock In {\em ACL}.

\bibitem[Frenkel et~al., 2021]{frenkel2021learning}
Frenkel, C., Lefebvre, M., and Bol, D. (2021).
\newblock Learning without feedback: Fixed random learning signals allow for
  feedforward training of deep neural networks.
\newblock {\em Frontiers in neuroscience}, 15.

\bibitem[Gilding and Kersner, 2004]{gilding}
Gilding, B.~H. and Kersner, R. (2004).
\newblock {\em Travelling Waves in Nonlinear Diffusion-Convection Reaction}.
\newblock Birkh\"{a}user.

\bibitem[Gong and Van~Leeuwen, 2009]{gong2009distributed}
Gong, P. and Van~Leeuwen, C. (2009).
\newblock Distributed dynamical computation in neural circuits with propagating
  coherent activity patterns.
\newblock {\em PLoS Computational Biology}, 5(12):e1000611.

\bibitem[Gregor and LeCun, 2010]{gregor2010emergence}
Gregor, K. and LeCun, Y. (2010).
\newblock Emergence of complex-like cells in a temporal product network with
  local receptive fields.

\bibitem[Gribizis et~al., 2019]{gribizis}
Gribizis, A., Ge, X., Daigle, T.~L., Ackman, J.~B., Zeng, H., Lee, D., and
  Crair, M.~C. (2019).
\newblock Visual cortex gains independence from peripheral drive before eye
  opening.
\newblock {\em Neuron}, 104(4):711--723.

\bibitem[Han et~al., 2008]{han2008reverberation}
Han, F., Caporale, N., and Dan, Y. (2008).
\newblock Reverberation of recent visual experience in spontaneous cortical
  waves.
\newblock {\em Neuron}, 60(2):321--327.

\bibitem[Hochreiter and Schmidhuber, 1997]{hochreiter1997long}
Hochreiter, S. and Schmidhuber, J. (1997).
\newblock Long short-term memory.
\newblock {\em Neural computation}, 9(8):1735--1780.

\bibitem[Huber et~al., 2020]{huber}
Huber, L., Finn, E., Handwerker, D., Boenstrup, M., Glen, D., Kashyap, S.,
  Ivanov, D., Petridou, N., Marrett, S., Goense, J., Poser, B., and Bandettini,
  P. (2020).
\newblock Sub-millimeter fmri reveals multiple topographical digit
  representations that form action maps in human motor cortex.
\newblock {\em NeuroImage}, 208.

\bibitem[Huth et~al., 2016]{huth}
Huth, A., {De Heer}, W., Griffiths, T., Theunissen, F., and Gallant, J. (2016).
\newblock Natural speech reveals the semantic maps that tile human cerebral
  cortex.
\newblock {\em Nature}, 532(7600):453--458.

\bibitem[Hyv\"{a}rinen and Hoyer, 2001]{hoyer}
Hyv\"{a}rinen, A. and Hoyer, P.~O. (2001).
\newblock A two-layer sparse coding model learns simple and complex cell
  receptive fields and topography from natural images.
\newblock {\em Vision Research}, 41(18):2413--2423.

\bibitem[Isaacson and Scanziani, 2011]{isaacson:2011}
Isaacson, J.~S. and Scanziani, M. (2011).
\newblock How inhibition shapes cortical activity.
\newblock {\em Neuron}, 72(2):231--243.

\bibitem[Issa et~al., 2000]{issa}
Issa, N.~P., Trepel, C., and Stryker, M.~P. (2000).
\newblock Spatial frequency maps in cat visual cortex.
\newblock {\em Journal of Neuroscience}, 20(22):8504--8514.

\bibitem[Keane and Gong, 2015]{keane}
Keane, A. and Gong, P. (2015).
\newblock Propagating waves can explain irregular neural dynamics.
\newblock {\em Journal of Neuroscience}, 35(4):1591--1605.

\bibitem[Launay et~al., 2020]{launay2020direct}
Launay, J., Poli, I., Boniface, F., and Krzakala, F. (2020).
\newblock Direct feedback alignment scales to modern deep learning tasks and
  architectures.
\newblock {\em arXiv preprint arXiv:2006.12878}.

\bibitem[Lillicrap et~al., 2016]{lillicrap2016random}
Lillicrap, T.~P., Cownden, D., Tweed, D.~B., and Akerman, C.~J. (2016).
\newblock Random synaptic feedback weights support error backpropagation for
  deep learning.
\newblock {\em Nature communications}, 7(1):1--10.

\bibitem[Makin et~al., 2015]{makin}
Makin, T.~R., Scholz, J., Henderson~Slater, D., Johansen-Berg, H., and Tracey,
  I. (2015).
\newblock {Reassessing cortical reorganization in the primary sensorimotor
  cortex following arm amputation}.
\newblock {\em Brain}, 138(8):2140--2146.

\bibitem[Muller et~al., 2014]{muller2014stimulus}
Muller, L., Reynaud, A., Chavane, F., and Destexhe, A. (2014).
\newblock The stimulus-evoked population response in visual cortex of awake
  monkey is a propagating wave.
\newblock {\em Nature communications}, 5(1):1--14.

\bibitem[N{\o}kland, 2016]{nokland2016direct}
N{\o}kland, A. (2016).
\newblock Direct feedback alignment provides learning in deep neural networks.
\newblock {\em arXiv preprint arXiv:1609.01596}.

\bibitem[Oja, 1982]{oja}
Oja, E. (1982).
\newblock A simplified neuron model as a principal component analyzer.
\newblock {\em J. Mathematical Biology}, 15:267--273.

\bibitem[Olshausen and Field, 1996]{Olshausen:Field:96}
Olshausen, B.~A. and Field, D.~J. (1996).
\newblock Emergence of simple-cell receptive field properties by learning a
  sparse code for natural images.
\newblock {\em Nature}, 381:607--609.

\bibitem[Osan and Ermentrout, 2001]{osan2001two}
Osan, R. and Ermentrout, B. (2001).
\newblock Two dimensional synaptically generated traveling waves in a
  theta-neuron neural network.
\newblock {\em Neurocomputing}, 38:789--795.

\bibitem[Ott et~al., 2019]{ott2019fairseq}
Ott, M., Edunov, S., Baevski, A., Fan, A., Gross, S., Ng, N., Grangier, D., and
  Auli, M. (2019).
\newblock fairseq: A fast, extensible toolkit for sequence modeling.

\bibitem[Pennington et~al., 2014]{pennington2014glove}
Pennington, J., Socher, R., and Manning, C.~D. (2014).
\newblock Glove: Global vectors for word representation.
\newblock In {\em Proceedings of the 2014 conference on empirical methods in
  natural language processing (EMNLP)}, pages 1532--1543.

\bibitem[Rozell et~al., 2008]{rozell:08}
Rozell, C.~J., Johnson, D.~H., Baraniuk, R.~G., and Olshausen, B.~A. (2008).
\newblock Sparse coding via thresholding and local competition in neural
  circuits.
\newblock {\em Neural Comput.}, 20(10):2526--2563.

\bibitem[Rubin et~al., 2017]{rubin}
Rubin, R., Abbott, L.~F., and Sompolinsky, H. (2017).
\newblock Balanced excitation and inhibition are required for high-capacity,
  noise-robust neuronal selectivity.
\newblock {\em Proc. Nat. Acac. Science}, 114(44):9366--9375.

\bibitem[Rubino et~al., 2006]{rubino2006propagating}
Rubino, D., Robbins, K.~A., and Hatsopoulos, N.~G. (2006).
\newblock Propagating waves mediate information transfer in the motor cortex.
\newblock {\em Nature neuroscience}, 9(12):1549--1557.

\bibitem[Santos et~al., 2014]{santos2014radial}
Santos, E., Sch{\"o}ll, M., S{\'a}nchez-Porras, R., Dahlem, M.~A., Silos, H.,
  Unterberg, A., Dickhaus, H., and Sakowitz, O.~W. (2014).
\newblock Radial, spiral and reverberating waves of spreading depolarization
  occur in the gyrencephalic brain.
\newblock {\em Neuroimage}, 99:244--255.

\bibitem[Sato et~al., 2012]{sato2012traveling}
Sato, T.~K., Nauhaus, I., and Carandini, M. (2012).
\newblock Traveling waves in visual cortex.
\newblock {\em Neuron}, 75(2):218--229.

\bibitem[Senk et~al., 2020]{senk2020conditions}
Senk, J., Korvasov{\'a}, K., Schuecker, J., Hagen, E., Tetzlaff, T., Diesmann,
  M., and Helias, M. (2020).
\newblock Conditions for wave trains in spiking neural networks.
\newblock {\em Physical review research}, 2(2):023174.

\bibitem[Shadlen and Newsome, 1998]{shadlen}
Shadlen, M.~N. and Newsome, W.~T. (1998).
\newblock The variable discharge of cortical neurons: {I}mplications for
  connectivity, computation, and information coding.
\newblock {\em Journal of Neuroscience}, 18:3870--3896.

\bibitem[Song et~al., 2021]{song}
Song, G., Xu, R., and Lafferty, J. (2021).
\newblock Convergence and alignment of gradient descent with random
  backpropagation weights.
\newblock {\em CoRR}, abs/2106.06044.

\bibitem[Tanaka et~al., 2019]{ganguli19}
Tanaka, H., Nayebi, A., Maheswaranathan, N., McIntosh, L., Baccus, S.~A., and
  Ganguli, S. (2019).
\newblock From deep learning to mechanistic understanding in neuroscience:
  {T}he structure of retinal prediction.
\newblock In {\em Advances in Neural Information Processing Systems}.

\bibitem[Templeton, 2021]{templeton}
Templeton, A. (2021).
\newblock Word equations: Inherently interpretable sparse word embeddings
  through sparse coding.
\newblock {\em Proceedings of the Fourth BlackboxNLP Workshop on Analyzing and
  Interpreting Neural Networks for NLP}.

\bibitem[Turing, 1952]{turing}
Turing, A. (1952).
\newblock The chemical basis of morphogenesis.
\newblock {\em Phil. Trans. Royal Society of London}, 237(641):37--72.

\bibitem[van~der Maaten and Hinton, 2008]{vanDerMaaten2008}
van~der Maaten, L. and Hinton, G. (2008).
\newblock Visualizing data using {t-SNE}.
\newblock {\em Journal of Machine Learning Research}, 9:2579--2605.

\bibitem[Wesselink et~al., 2020]{wesselink}
Wesselink, D.~B., Sanders, Z.-B., Edmondson, L.~R., Dempsey-Jones, H., Kieliba,
  P., Kikkert, S., Themistocleous, A.~C., Emir, U., Diedrichsen, J., Saal,
  H.~P., and Makin, T.~R. (2020).
\newblock Malleability of the cortical hand map following a finger nerve block.
\newblock {\em bioRxiv}.

\bibitem[Wooley et~al., 2017]{wooley}
Wooley, T.~E., Baker, R.~E., and Maini, P.~K. (2017).
\newblock Turing's theory of morphogenesis.
\newblock In Copeland, B.~J., Bowen, J.~P., Wilson, R., and Sprevak, M.,
  editors, {\em The Turing Guide}. Oxford University Press.

\bibitem[Yogatama et~al., 2015]{yogatama}
Yogatama, D., Faruqui, M., Dyer, C., and Smith, N. (2015).
\newblock Learning word representations with hierarchical sparse coding.
\newblock In Bach, F. and Blei, D., editors, {\em Proceedings of the 32nd
  International Conference on Machine Learning}, volume~37 of {\em Proceedings
  of Machine Learning Research}, pages 87--96, Lille, France. PMLR.

\bibitem[Zanos et~al., 2015]{zanos2015sensorimotor}
Zanos, T.~P., Mineault, P.~J., Nasiotis, K.~T., Guitton, D., and Pack, C.~C.
  (2015).
\newblock A sensorimotor role for traveling waves in primate visual cortex.
\newblock {\em Neuron}, 85(3):615--627.

\bibitem[Zhou et~al., 2021]{zhou21}
Zhou, B., Li, Z., Kim, S. S.~Y., Lafferty, J., and Clark, D.~A. (2021).
\newblock Shallow neural networks trained to detect collisions recover features
  of visual loom-selective neurons.
\newblock {\em bioRxiv}.

\bibitem[Zylberberg et~al., 2011]{zylberberg}
Zylberberg, J., Murphy, J.~T., and DeWeese, M.~R. (2011).
\newblock A sparse coding model with synaptically local plasticity and spiking
  neurons can account for the diverse shapes of {V1} simple cell receptive
  fields.
\newblock {\em PLOS Computational Biology}, 7:1--12.

\end{thebibliography}

\end{document}